\documentclass[prb,aps,twocolumn,showpacs]{revtex4}

\usepackage{amssymb}
\usepackage{amsmath}
\usepackage{graphicx}
\usepackage{color}

\begin{document}

\title{Polariton propagation in weak confinement quantum wells.}
\author{D. Schiumarini, N. Tomassini, L. Pilozzi, A. D'Andrea}
\address{Istituto dei Sistemi Complessi, CNR, C.P. 10, 
Monterotondo Stazione, Roma I-00016}
\date{\today }

\begin{abstract}
Exciton-polariton propagation in a quantum well, under centre-of-mass quantization, is computed by a variational self-consistent microscopic theory. The Wannier exciton envelope functions basis set is given by the simple analytical model of ref. [1], based on pure states of the centre-of-mass wave vector, free from fitting parameters and "ad hoc" (the so called additional boundary conditions-ABCs) assumptions. In the present paper, the former analytical model is implemented in order to reproduce the centre-of-mass quantization in a large range of quantum well thicknesses ($5a_B \leq L \leq \infty $). The role of the dynamical transition layer at the well/barrier interfaces is discussed at variance of the classical Pekar's dead-layer and ABCs. The Wannier exciton eigenstates are computed, and compared with various theoretical models with different degrees of accuracy. Exciton-polariton transmission spectra in large quantum wells ($L\gg a_B$) are computed and compared with experimental results of Schneider et al.\cite{Schneider} in high quality GaAs samples. The sound agreement between theory and experiment allows to unambiguously assign the exciton-polariton dips of the transmission spectrum to the pure states of the Wannier exciton center-of-mass quantization.

\end{abstract}


\maketitle
\textit{In memory of Prof. Dr. Giuseppe Franco Bassani (1929-2008)}

\section{Introduction}
Exciton-polariton propagation in mesoscopic super-structures has shown new interesting physical phenomena. Super-radiant effects in multi-quantum wells\cite{Ivchenko,Stroucken,Hubner1,pilozzi}, polariton propagation in uniaxial mesoscopic crystals\cite{Tomassini}, soliton-polariton and shock waves propagation\cite{Egorov}, polaritons condensation in DBR cavities\cite{amo}, are present in the recent literature. All these phenomena are usually observed in exciton-polariton propagation experiments performed in optically non-local meso-structures. Among these, DBR micro-cavities and multi-quantum wells under Bragg condition are usually studied as paradigmatic systems showing strong radiation-matter interaction\cite{Weisbuch,Hubner,Schaefer,Huang}. For instance, periodic N-quantum wells are usually well suited for studying super-radiant exciton-polariton propagation, that crucially depends on the optical properties of theirs 2N-interfaces\cite{pilozzi,Tomassini}. Moreover, if no-overlapping between the exciton envelope functions, localized in adjacent quantum wells, are present, the so called dead-layer model was largely used in order to describe the optical properties of the 2N-interfaces of the system \cite{Goldberg,Halevi,Tredicucci,Tuffigo,Tomassini2,Henneberger,Muljiarov,Schumacher1}. Recently, a large dead layer effect was also observed in colloidal ZnO nano crystals, and this surface optical passivation should be responsible on shorten the exciton lifetime in this kind of systems.\cite{Vladimir} 

In his pioneering paper\cite{Pekar}, Pekar introduced the concept of exciton "dead-layer" (or extrinsic dead-layer) in order to justify the three-layers model of the electric dipole polarization necessary to reproduce the reflectance lineshape of Wannier exciton spectra in a semi-infinite semiconductor. In a subsequent paper\cite{Hopfield}, Thomas and Hopfield discussed the microscopic basis (interface structural disorder, chemical disorder, image potential, impurities etc\textellipsis) of this classical effect.  Looking forward a self-consistent microscopic computation of polariton propagation in semi-infinite samples in the semiclassical framework, D'Andrea and Del Sole\cite{da15} introduced an "intrinsic dead layer" (or transition layer) by imposing the minimization of the exciton envelope function mismatch at the vacuum/semiconductor surface, expanded in an hydrogenic basis set. 
In the same paper an analytical approximation of the semi-infinite exciton envelope function, that exactly accomplishes the so called no-escape boundary conditions (NEBCs) at surface plane (Z=0) of the sample, namely: $\Psi _K (z_e  = 0) = \Psi _K (z_h  = 0) = 0$, was also given for computing the optical response at semiconductor band edge\cite{da15,da16}:
\begin{eqnarray}
\Psi _K (\vec r,\vec R) =N_K \left[ {e^{ - iKZ}  + A\,e^{iKZ}  - \left( {1 + A} \right)e^{ - \bar PZ} } \right]\\
\varphi _{1s} (r)\,e^{i\vec K_{//}  \cdot \vec R_{//} } /\sqrt S\nonumber
\end{eqnarray}
where $\vec r$ and $\vec R$ are the coordinates of the electron-hole relative and center-of-mass motion respectively, $\varphi _{1s} (r)$
is the n=1 hydrogenic wave function of the relative motion, K is the wave vector of the centre-of-mass along Z-axis (with $Z\geq 0$), $\vec K_{//} $ is the corresponding in-plane wave vector, $A =  - \frac{{\bar P - iK}}{{\bar P + iK}}$ is the exciton reflection amplitude on Z=0 vacuum/semiconductor surface ($\left| A \right|^2 =1$) and $\bar P$ is an average coefficient of the evanescent waves due to the higher energy hydrogenic wave functions ($n>1$) taken close to the continuum states ($n \to \infty $)\cite{da15,da16}. Moreover the exciton energy is: $E_K  =  - R^* (a_B ) + \frac{{\hbar ^2 }}{{2M}}\left[ {K^2  + K_{//}^2 } \right]$ where $R^* (a_B )$ is the 3D Rydberg, M and $\mu$ are total and reduced mass respectively, and $a_B $ is the 3D Bohr radius of the exciton. The transition layer $1/\bar P = \delta _\infty  $
 is strongly dependent on the center-of-mass wave vector value K along Z-axis [25], and coincides with the zone, close to the surface plane Z=0, where the exciton envelope function $\Psi _K (z_e ;z_h ;\vec r) \approx 0$\cite{da15,da16}.

Notice that the transition layer of eq.(1) is a quantum  mechanical quantity derived self-consistently by a variational principle, and due to the distortion of the exciton envelope function at surface, where centre-of-mass and relative motion of  the exciton are entangled, while Pekar's dead-layer is a classical quantity characterizing the zone of the surface where the electric polarization is zero (additional boundary condition-ABC), and due to the selvedge layer present in real samples\cite{Hopfield}. Therefore, since Pekar's dead-layer is usually determined as fitting parameter, its value can take into account not only the effect due to the boundary conditions imposed on the exciton envelope function (intrinsic dead-layer), but also other physical effects present in real surfaces\cite{Hopfield} (extrinsic dead-layer).

The former analytical model of eq.(1) was extended to the slab geometry\cite{dads1} where the transition layer and the effective Bohr radius were determined by a variational minimization of the first momentum of the exciton Hamiltonian. Also in this case the average level was chosen close to the energy of the hydrogenic continuum states ($n\to \infty $)\cite{dads1}. Notice, that the former choice, not strictly necessary for the consistency of the theory, induces some inconsistency as pointed out by the authors of ref.[27,28]. In the present paper a different choice, not affected by the former problem and that nicely reconciles\cite{19} the transition layer in semi-infinite solids and in slabs, is suggested.

Exciton envelope functions in quantum confined systems (wells, wires and dots) are usually computed by variational minimization of the exciton energy in a so called "ABC free" theory\cite{Ishihara}.

Notice, that in the large quantum well limit ($L \gg a_B$) a very accurate numerical minimization is necessary in order to determine a sensible dead-layer value. In fact, we remind that the interband transitions at semiconductor band edge are of the order of some electronvolts in the visible range of energies, and the Wannier exciton energies are close to the hydrogenic Rydberg: $
R^* (a_B ) = \frac{{\hbar ^2 }}{{2\mu }}\frac{1}{{a_B^2 }}$ (of the order of tens of meV), while the centre-of-mass energy of the exciton is given by: $E_{CM} (L \gg a_B ) \approx \frac{{\hbar ^2 }}{{2M}}\left( {\frac{\pi }{L}} \right)^2$ (of the order of tens of $\mu $eV). Finally, the sensitiveness of the optical spectrum to the dead-layer effect can be given by the centre-of-mass energy difference with and without the dead-layer effect, namely: $
\Delta E(d) = E(L - 2d) - E(L) \approx \frac{{\hbar ^2 }}{{2M}}\left( {\frac{\pi }{L}} \right)^2 \frac{{4d}}{L}$ (from about tens of $\mu eV$ to zero), where $\Delta E(d)/E(L) \approx 4d/L$. Therefore, many orders of magnitude\cite{schiumarini} in the energy accuracy should be necessary for computing the Wannier exciton dynamics under the centre-of-mass quantization.
 
Notice, that the transition layer affects the exciton centre-of-mass momentum (spatial dispersion effect), and gives additional light waves propagating into the sample, therefore its effect on the optical properties cannot be neglected neither in the presence of a large energy broadening. Even if it has a rather negligible effect on the energy position of exciton-polariton peaks in transmission spectra, we expect that it should have a non negligible effect on the line shape analysis, and its role should be even enhanced in the time resolved optical spectra in multi-quantum well systems\cite{Egorov,amo}.

The aim of the present work is two-folds: i) a non-adiabatic Wannier exciton envelope function, obtained as a generalization of the analytical model of ref.[1], is solved by a variational method, and compared with other model calculations at different degree of accuracy, namely: a more accurate variational expansion in e-h subband products\cite{atanasov}, an exact adiabatic solution\cite{Combescot18} and the heuristic so called "hard sphere" model. Notice, that the former extension is necessary in order to study the limit of validity of the analytical model itself, and it becomes also mandatory when a complete one exciton basis set must be used, as in non-linear optical computation\cite{glazov}.
ii) In order to check the ability of the former analytical model in reproducing the exciton centre-of-mass dynamics, the self-consistent optical transmission spectra in a single quantum well is computed and compared with experimental results of Schneider et al.\cite{Schneider} performed in high quality GaAs quantum wells. A sound agreement is obtained for high quality quantum wells both for intensity and phase measurements of the optical transmission spectrum with our analytical model for exciton masses ratio from the positronium limit ($\mu /M=0.248$) to the hydrogenic one ($\mu /M=0.111$); moreover, it allows to unambiguously assign the higher energy peaks, present in the spectrum of ref.[2], to the pure states of the centre-of-mass quantization.

In Sect. II the analytical Wannier exciton function model of ref.[1] is revisited and generalized. We here remind all the approximations involved in order to facilitate the generalization of the analytical model at an order higher than the 1s hydrogenic wave function as explicitly reported in the appendix A.

In Sect. III the optical response of a single quantum well is computed and compared with the experimental results of Schneider et al.\cite{Schneider}.

\section{Wannier exciton in a slab: variational analytical model revisited and implemented.}

Let us consider a Wannier exciton of energy E, confined in a single quantum well of thickness $L \gg a_B$, and clad between two infinite potential barriers ($ - L/2 \leq Z \leq L/2$). The exciton function, in a two band model and in effective mass approximation, can be expanded in a complete basis set of hydrogenic eigenstates\cite{dads1}: $\hat H_{\vec r} \,\varphi _{nlm} (\vec r) = \varepsilon _n \,\varphi _{nlm} (\vec r)$ , where $\hat H_{\vec r}  =  - \frac{{\hbar ^2 }}{{2\mu }}\vec \nabla _{\vec r}^2  - \frac{{e^2 }}{{\varepsilon _b r}}$,
\begin{subequations}
\begin{align}
\Psi _K (\vec r,\vec R) =\Psi _E (\vec r,Z)\,e^{i\vec K_{//} \cdot \vec R_{//} } /\sqrt S\hspace{1.5in}\\
\Psi _E (\vec r,Z)=\sum\limits_{nlm} {\left[ {a_{nlm} e^{iK_n Z}  + b_{nlm} e^{ - iK_n Z} } \right]}\varphi _{nlm} (\vec r)\hspace{0.6in}
\end{align}
\end{subequations} 
The modulus of the exciton centre-of-mass wave vector along Z-axis is: $
K_n =\sqrt {K^2 - K_{//}^2} 
  =\sqrt { \left[ {\frac{{2M}}
{{\hbar ^2 }}\left( {E - \varepsilon _n } \right)} \right] - K_{//}^2} $ where $
\vec K_{//}$ is the wave vector along the (x,y) plane.
Now, the sum on the hydrogenic basis set of eq.(2) can be separated in a finite ( for $E \geq \varepsilon _n  + \hbar ^2 K_{//}^2 /2M$)  and an infinite sum (with $E < \varepsilon _n  + \hbar ^2 K_{//}^2 /2M$), that, at normal incidence configuration ($K_{//}  = 0.0$), give the following exciton envelope function: 
\begin{eqnarray}
\Psi_E (\vec r,Z) =\hspace{2.4in}\\
=\sum\limits_{nlm}^{\left\{ {E \geq \varepsilon _n } \right\}} {\left[ {\alpha _{nlm} \cos \left( {K_n Z} \right) + \beta _{nlm} \sin \left( {K_n Z} \right)} \right]\,\varphi _{nlm} (\vec r)}\nonumber\\
+ \sum\limits_{nlm}^{\left\{ {\varepsilon _n  > E} \right\}} {\left[ {a_{nlm} e^{ - P_n Z}  + b_{nlm} e^{P_n Z} } \right]\,\varphi _{nlm} (\vec r)}\nonumber
\end{eqnarray} 
where $P_n  = iK_n  = \left[ {\frac{{2M}}
{{\hbar ^2 }}\left( {\varepsilon _n  - E} \right)} \right]^{1/2} $ and $\alpha _{nlm}  = a_{nlm}  + b_{nlm}$,\, $\beta _{nlm}  = i\left( {a_{nlm}  - b_{nlm} } \right)$.

The analytical approximation of the former wave function is obtained by adopting the procedure of ref. [31]. First of all, we restrict the finite sum to only one hydrogenic function: the 1s (n=1) state; the general case will be discussed in appendix A and specialized for n=2.

The computation for a symmetric quantum well proceeds by defining the operator: $\hat{\mathbb{R}}(z \to  - z;Z \to  - Z)$ that, applied to the exciton envelope function of eq.(3), gives the relationship: $\hat{\mathbb{R}}\;\Psi _E (\vec r,Z) =  \pm \Psi _E (\vec r,Z)$. From this relation two sets of exciton envelope functions are generated for even and odd symmetry respectively: 
\begin{subequations}
\begin{align}
\Psi _{_E }^e (\vec r,Z) = \alpha _1 \cos \left( {K_1 Z} \right)\varphi _{_{1s} }^{} (r) +\hspace{1.5in} \\
2\sum\limits_{nlm}^{n > 1} {\left[ {\alpha _{nlm} \cosh \left( {P_n Z} \right)\varphi _{_{nlm} }^e (\vec r) - \beta _{nlm} \sinh \left( {P_n Z} \right)\varphi _{_{nlm} }^o (\vec r)} \right]}\nonumber\\
\nonumber\\  
\Psi _{_E }^o (\vec r,Z)\, = \beta _1 \sin \left( {K_1 Z} \right)\varphi _{_{1s} }^{} (r) +\hspace{1.5in} \\
2\sum\limits_{nlm}^{n > 1} {\left[ {\alpha _{nlm} \cosh \left( {P_n Z} \right)\varphi _{_{nlm} }^o (\vec r) - \beta _{nlm} \sinh \left( {P_n Z} \right)\varphi _{_{nlm} }^e (\vec r)} \right]}\nonumber
\end{align}  
\end{subequations}
where the coefficients $\alpha _1$ and $\beta _1 $ are: $\alpha _1 = \alpha _{100}$ and $\beta _1 = \beta _{100}$.

In order to perform a smart truncation of the former series expansion, let us consider N-hydrogenic functions (with $N \to \infty$), and substitute the infinite sum of virtual states with an (N-1)-times degenerate state, located at an energy value much higher than E. Notice that the former approximation can be obtained by substituting $P_{n > 1} $ with an average value $ \bar P = 1/\delta $, where $\delta$ is the transition layer of the Wannier exciton\cite{dads1}. Therefore the former "one transition layer" approximation of Wannier exciton envelope function is based on the substitution of the infinite sum of the hydrogenic basis set in eqs. (4a) and (4b) with four analytical terminators that accomplish the no-escape boundary conditions (NEBCs) at the well/barrier interfaces. In fact, by imposing the NEBC: $\Psi _E (z_e  =  \pm L/2,z_h ,\vec \rho ) = \Psi _E (z_h  =  \pm L/2,z_e ,\vec \rho ) = 0.0$ we can determine, as shown in refs. [1,31], the following four rational functions:
\begin{eqnarray*}
F_{ee} (\vec r) = 2\,\sum\limits_{nlm}^{n > 1} {(\alpha _{nlm} /\alpha _1 )\,\varphi _{_{nlm} }^e (\vec r)}  = f_{ee} (z)\,\varphi _{1s} (r)\\
F_{oo} (\vec r) = 2\,\sum\limits_{nlm}^{n > 1} {(\alpha _{nlm} /\alpha _1 )\,\varphi _{_{nlm} }^o (\vec r)}  = f_{oo} (z)\,\varphi _{1s} (r)\\
F_{eo} (\vec r) = 2\,\sum\limits_{nlm}^{n > 1} {(\beta _{nlm} /\beta _1 )\,\varphi _{_{nlm} }^o (\vec r)}  = f_{eo} (z)\,\varphi _{1s} (r)\\
F_{oe} (\vec r) = 2\,\sum\limits_{nlm}^{n > 1} {(\beta _{nlm} /\beta _1 )\,\varphi _{_{nlm} }^e (\vec r)}  = f_{oe} (z)\,\varphi _{1s} (r)
\end{eqnarray*}
that give the requested terminators as shown explicitly in refs. [1,31] (and also in the appendix A see eqs. (5a)-(5)b).
This approximation gives rather accurate numerical results when a large energy gap is present between the travelling 1s state and the virtual higher energy states, as will be carefully checked in the calculation and further discussed in this section.
With the use of the former terminators the exciton envelope functions assume a compact analytical form:
\begin{subequations}
\begin{align}
\Psi _m^e (\vec r,Z)=N_m^e[{\cos (K_m Z)\, + \cosh (\bar P\,Z)\,f_{_{ee} }^{(m)} (z)}\\
{ - \sinh (\bar PZ)\,f_{_{eo} }^{(m)} (z)}]\varphi _{1s} (r) =\nonumber\\
=N_m^e \,\,g_m^e (z,Z)\;\varphi _{1s} (r) \;for \;m=1,3,5,\textellipsis \nonumber\\
\nonumber\\
\Psi _m^o (\vec r,Z) = N_m^o [{\sin (K_m Z)\, + \cosh (\bar PZ)\,f_{_{oo} }^{(m)} (z)}\\
{ - \sinh (\bar P\,Z)\,f_{_{oe} }^{(m)} (z)}]\varphi _{1s} (r) =\nonumber\\
= N_m^o \,\,g_m^o (z,Z)\;\varphi _{1s} (r) \;for \;m=2,4,6,\textellipsis \nonumber
\end{align}
\end{subequations}
where $K_m  \equiv K_1 (m)$ with m the centre-of-mass quantum number for even an odd symmetry, and $\varphi _{1s} (r) = e^{ - r/a} /\sqrt {\pi a^3 }$ is the 1s hydrogenic function of the relative motion.

Finally, from the continuity of the exciton wave function $\Psi ^{(i)} (z \to 0^ +  ) = \Psi ^{(i)} (z \to 0^ -  )$
 and its first derivative $
\left. {\frac{{\partial \Psi ^{(i)} }}
{{\partial z}}} \right|_{z = 0^ +  }  = \left. {\frac{{\partial \Psi ^{(i)} }}
{{\partial z}}} \right|_{z = 0^ -  }$ for i=e,o at z=0 surface of the relative motion, the relationships for the center-of-mass quantization are obtained. Notice, that, due to the symmetry properties of the exciton envelope functions, the two former relations are reduced to the condition of approaching the z-surface with zero first order derivative, namely: $\left. {\frac{{\partial F_{_{ee} }^{} }}{{\partial z}}} \right|_{z = 0}  = 0$ and $\left. {\frac{{\partial F_{_{oe} }^{} }}{{\partial z}}} \right|_{z = 0}  = 0$. These conditions give the centre-of-mass dispersion relations for even (m=1,3,5\textellipsis) and odd (m=2,4,6\textellipsis) exciton respectively, namely:
\begin{subequations}
\begin{align}
K_m \,tg\left[ {K_m \,L/2} \right] + \bar P\,tgh\left[ {\bar P\,L/2} \right] = 0\;for\;m=1,3,5,\textellipsis\\
\bar P\,tg\left[ {K_m \,L/2} \right] - \,K_m \,tgh\left[ {\bar P\,L/2} \right] = 0\;for\;m=2,4,6,\textellipsis
\end{align}
\end{subequations}
Obviously, the difference between the former quantization conditions and the simple condition $K_m = \frac{\pi }{L}\,m$ for m=1,2,3,..  is due to the composed nature of the Wannier exciton and eq. (6a) and (6b) will recover the former simple condition in the limit value $m \to \infty $.
\begin{figure}[t]
\includegraphics[scale=0.36]{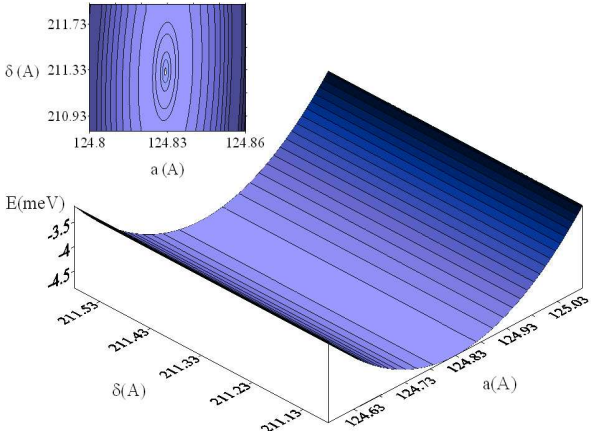}
\caption{(Color online) Exciton energy as a function of effective Bohr radius (a) and transition layer ($\delta$) parameters in a AlGaAs/GaAs(001) quantum well of tickness L=250nm. In the inset the minimum exciton energy is clearly shown.}
\end{figure}

The variational minimization of the first momentum of the exciton Hamiltonian is computed for the lowest even and odd energy states (m=1,2): $\left\langle {\Psi _m } \right|\hat H_{ex} \left| {\Psi _m } \right\rangle /\left\langle {\Psi _m } \right|\left| {\Psi _m } \right\rangle  = \min $, by taking as variational non-linear parameters the effective Bohr radius (a) and the inverse of the transition layer ($\bar P = 1/\delta $).  Even and odd variational parameters assume rather the same value for large quantum wells ($L \gg a_B $) and exactly the same value in the bulk limit ($L\to \infty$)\cite{dads1,da15}.

Since the dynamics of an exciton, perfectly confined in a quantum well, strongly depends on the well thickness (L/$a_B $), let us consider the following different zones of quantum well dimensions, namely: i) a zone of very large quantum wells ($L \gg a_B $) or bulk limit $L \to \infty $, where both the variational parameters assume their bulk values, and therefore the exciton energy converges to the "hard sphere model", namely: $E_m  =  - R^* (a_B ) + m^2 \frac{{\hbar ^2 }}{{2M}}\left( {\frac{\pi }{{L - 2d}}} \right)^2 $
for m=1,2,\textellipsis . 
In this zone the transition layer assumes its saturation value ($\delta  = \delta _{L \to \infty } $)\cite{schiumarini} in numerical agreement with that obtained in a semi-infinite sample\cite{viri,viri2}. ii) The centre-of-mass quantization zone ($10a_B  \leqslant L \leqslant 30a_B $), where the Bohr radius changes smoothly and its value remains rather close to the bulk one, while the transition layer variation strongly depends on the quantum well thickness and dominates the exciton dynamics from hydrogenic ($\mu /M$=0) to positronium behaviour ($\mu /M$=1/4). Notice, that in this zone the former analytival model is well suited for describing the exciton composite dynamics. Moreover, iii) by decreasing the quantum well thickness in the range $5a_B  \leqslant L \leqslant 10a_B $ also the Bohr radius value shrinks. In this zone strong non-adiabatic effects, due to the entanglement between centre-of-mass and relative motion, are present, and the analytic model could be implemented by higher energy hydrogenic wave function (see appendix A). iv) Finally, in the zone where Wannier exciton behaviour changes from  3D$ \to $2D dynamics, the transition layer and the centre-of-mass quantization along Z-axis loose their meanings.
 
In order to study the Wannier exciton behaviour in the centre-of-mass quantization the m=1 energy minimization is computed by adopting the formula reported in ref.[31]. In Fig.1 the exciton energy, in a rather large AlGaAs/GaAs(001) quantum well, is shown as a function of the variational parameters value . The physical parameter values of the model are the same of ref. [36], namely: $m_e  = $0.067$m_o $, $m_{hh}  = $0.457$m_o $, $n_b = 3.71$, and $L = 20a_B  = 250nm$. In this system the heavy hole Bohr radius and the Rydberg  energy are $a_B  = $12.4649nm and $R^* (a_B )$=-4.1965meV respectively. Therefore, the Wannier exciton is in the centre-of-mass quantization zone with a mass ratio $\mu /M = $ 0.111, rather close to the hydrogenic behaviour.
 
Notice, that in a scale where the minimum energy with respect to the effective Bohr radius is clearly shown, the one due to the transition layer is hardly observed (see the inset of Fig.1)\cite{25}. Moreover, the minimized effective exciton radius is a=12.4850$\pm$0.002nm, and the transition layer $\delta=1/\bar P = $21.160$\pm$0.002nm, while the exciton energy for m=1 envelope function is $\varepsilon _1 = $-4.17975meV, and the computed centre-of-mass wave vector along Z-axis is $K_{m = 1} = 0.795340 * 10^{ - 3} a.u.$
The wave vector and the exciton energy, derived from the minimization, are both in rather good agreement with those computed by the hard sphere model by taking the dead-layer value equal to the minimized transition layer ($d = \delta $).
\begin{figure}[t]
\includegraphics[scale=0.35]{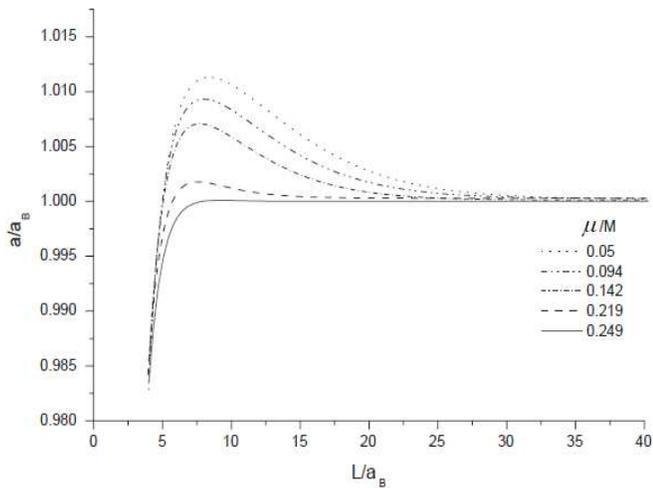}
\caption{Effective Bohr radius ($a/a_B $) as a function of quantum well thickness ($L/a_B $) computed for five different masses ratios ($\mu /M:$0.050; 0.094; 0.142; 0.219; 0.249).}
\end{figure}
\begin{figure}[b]
\includegraphics[scale=0.38]{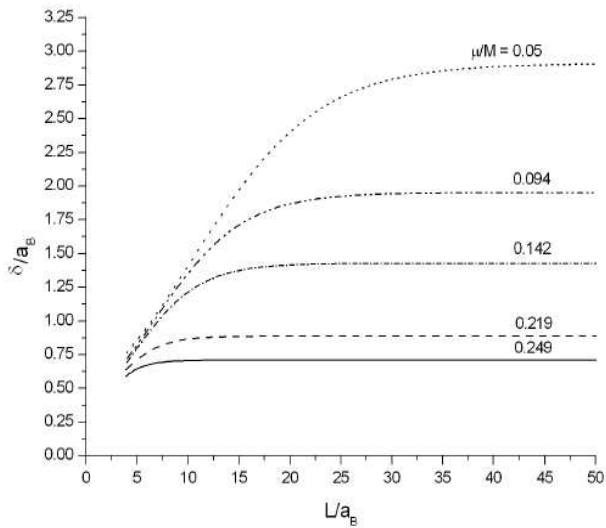}
\caption{Transition layer values ($\delta /a_B $) as a function of quantum well thickness ($L/a_B $) computed for five different masses ratios  ($\mu /M:$0.050; 0.094; 0.142; 0.219; 0.249).}
\end{figure}
In fact, the hard sphera model gives: $\tilde \varepsilon _1  =  - R^* (a_B ) + \hbar ^2 \tilde K_1^2 /2M =  - 4.17985meV$ and $\tilde K_1  = \pi /(L - 2\delta ) = 0.800525 * 10^{ - 3} a.u$, while exciton wave vector computed by zero dead layer value is strongly different ($\tilde K = \pi /L \approx 0.665012*10^{ - 3} a.u.$) from both the former values.

Notice, from eq.(6a) and (6b), that the minimized transition layer of the analytical model coincides with the dead-layer ($\delta  = d$) for m=1 center-of-mass wave vector, while increasing the m-value ($m > 1$) the dead layer value decreases ($\delta  > d$), till the limit value $m \to \infty $ where the centre-of-mass wave vector is $K_m  = m\,\pi /L$ in correspondence of negligible dead-layer value ($d\to 0$) (see also Fig. 5 and 6). 

In conclusion, the lowest energy exciton state (m=1) of our analytical model is in good agreement with the results obtained by the heuristic hard sphere model when the dead-layer is substituted by the corresponding minimized transition layer value. Moreover, the sound agreement with the "exact" adiabatic solution of ref. [29] is fully discussed in ref. [31], and will not be reported here again.

The transition layer of the non-adiabatic exciton envelope function of eqs. (5a) and (5b) introduces an entanglement between relative and centre-of-mass motion of the exciton, that is the very reason that makes needless the ABC imposed to the Maxwell equations, and that causes many electromagnetic waves propagation (spatial dispersion effect) also in semiconductor with cubic simmetry\cite{Pekar,Agranovich}.
\begin{figure}[t]
\includegraphics[scale=0.84]{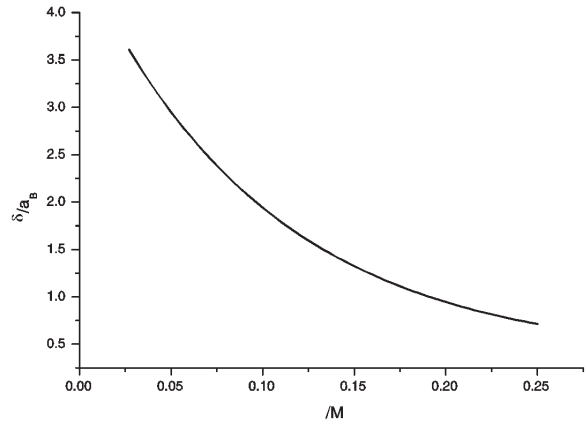}
\caption{Saturation value of the transition layer ($\delta _{L \to \infty } /a_B $) as a function of mass ratio $\mu /M$.}
\end{figure}

A systematic analysis of the effective Bohr radius and of the transition layer as a function of the quantum well thickness, and for different values of the mass ratio $\mu /M$ are shown in Fig.2 and Fig.3 respectively. The dependence of the effective Bohr radius on the well thickness (Fig.2) is a strong function of the mass ratio $\mu /M$. In fact, it reaches its saturation limit at about $10\,a_B $ for positronium, while the hydrogenic limit converges to the correct bulk value for well thicknesses larger than $30\,a_B $, and shows a small increment (less than 1/100) with respect to its bulk value at L$ \approx 7.5\,a_B $. Notice, that also the transition layer value reaches its saturation for L$ \approx 10\,a_B $ in the positronium, while well thicknesses as large as $30\,a_B $ are necessary in order to obtain the hydrogenic limit (Fig.3).

The asymptotic values of the transition layer for very large quantum wells ($L \to \infty $), as a function of mass ratio, are shown in Fig.4. We observe that the transition layer is as large as four-times the Bohr radius in the limit of hydrogenic masses ratio ($m_h  \approx 1836m_e  \Rightarrow \delta  \approx 4a_B $), while it is a bit greater than half the Bohr radius in the positronium limit ($M = 4\mu  \Rightarrow \delta  \approx 0.7a_B $); its behaviour is also in agreement with the exact adiabatic exciton envelope function\cite{Combescot18} as shown in a previous computation  for quantum well in the range of thicknesses $10a_B  \leqslant L \leqslant 20a_B $ (see Fig.3 ref. [31] where the two curves must be exchanged).
We would like to remind that a former evaluation\cite{Combescot18}, derived from the energy balance equation: $ - R^* (a) + \frac{{\hbar ^2 }}
{{2M}}K_m^2  =  - \frac{{\hbar ^2 }}
{{2M}}\left( {\frac{1}{\delta }} \right)^2 $, at the continuum limit of the hydrogenic set of states gives not a correct behaviour of the transition layer\cite{19} as a function of exciton masses ratio (see also ref.[31] and the discussion therein).

Notice that exciton transition layer of Fig.4, computed in the limit of slab thickness $L \to \infty $, should be equal to those in semi-infinite sample of eq.(1). Therefore, the former choice can reconcile the transition layer effect in semi-infinite samples\cite{viri,viri2} and in slabs\cite{dads1}, without introducing a further "ad hoc" dead layer (added to the transition layer) as hypothesized in ref.[34]. 

We have observed before, that in the analytical model of eqs. (5a) and (5b) the most intriguing approximation is the hydrogenic series truncation adopted. In the present paper, the correct energy location of the average virtual state\cite{Combescot18} is obtained by comparing the analytical model with the full theory of ref. [32].
\begin{figure}[t]
\includegraphics[scale=0.36]{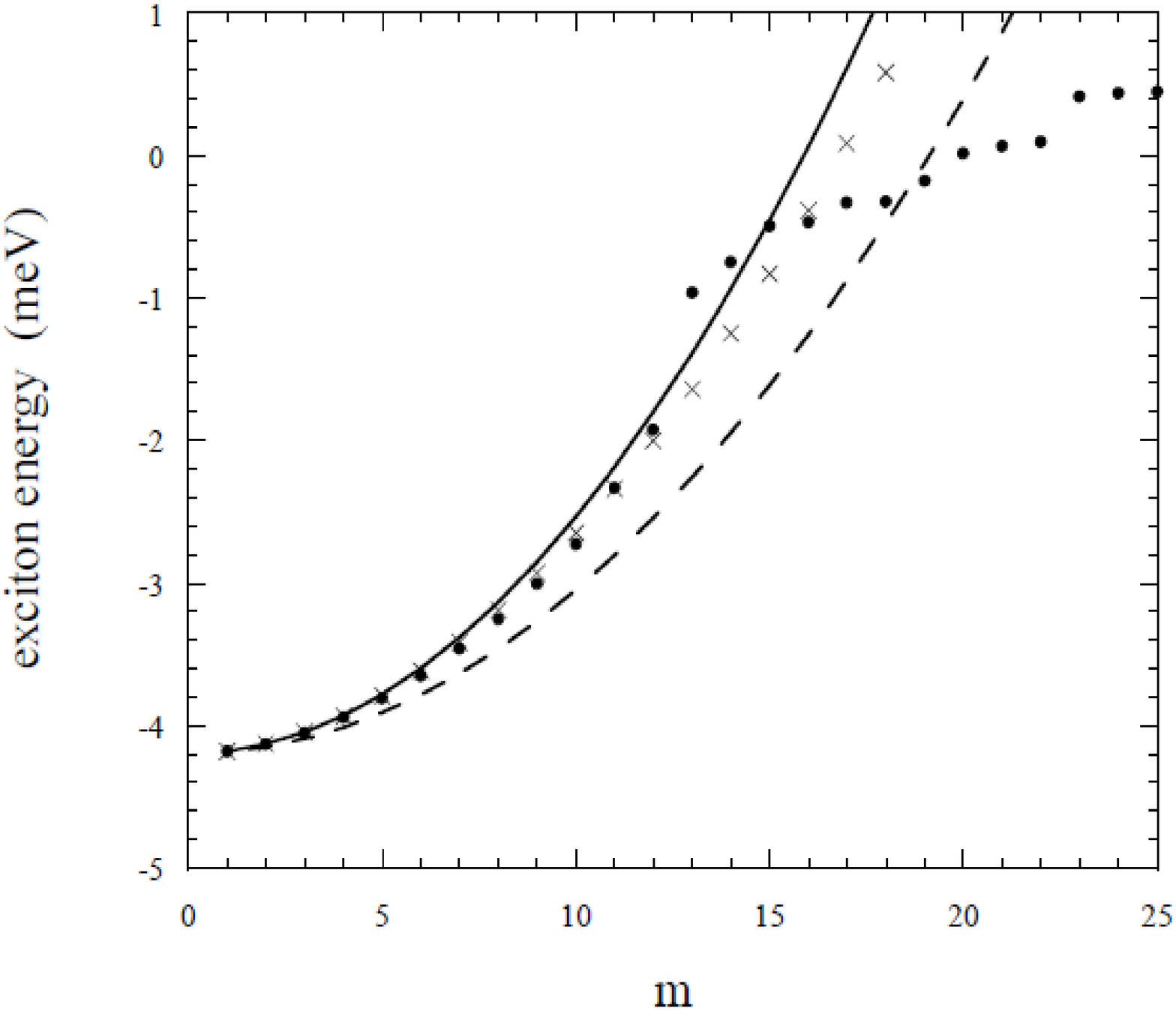}
\caption{Heavy hole exciton energy as a function of centre-of mass quantum number m in a AlGaAs/GaAs(001) quantum well of thickness L=250nm. The energy is computed by full theory (dots), analytical model (crosses), hard sphere model with (solid curve) and without dead layer (dashed curve).}
\end{figure}
\begin{figure}[b]
\includegraphics[scale=0.36]{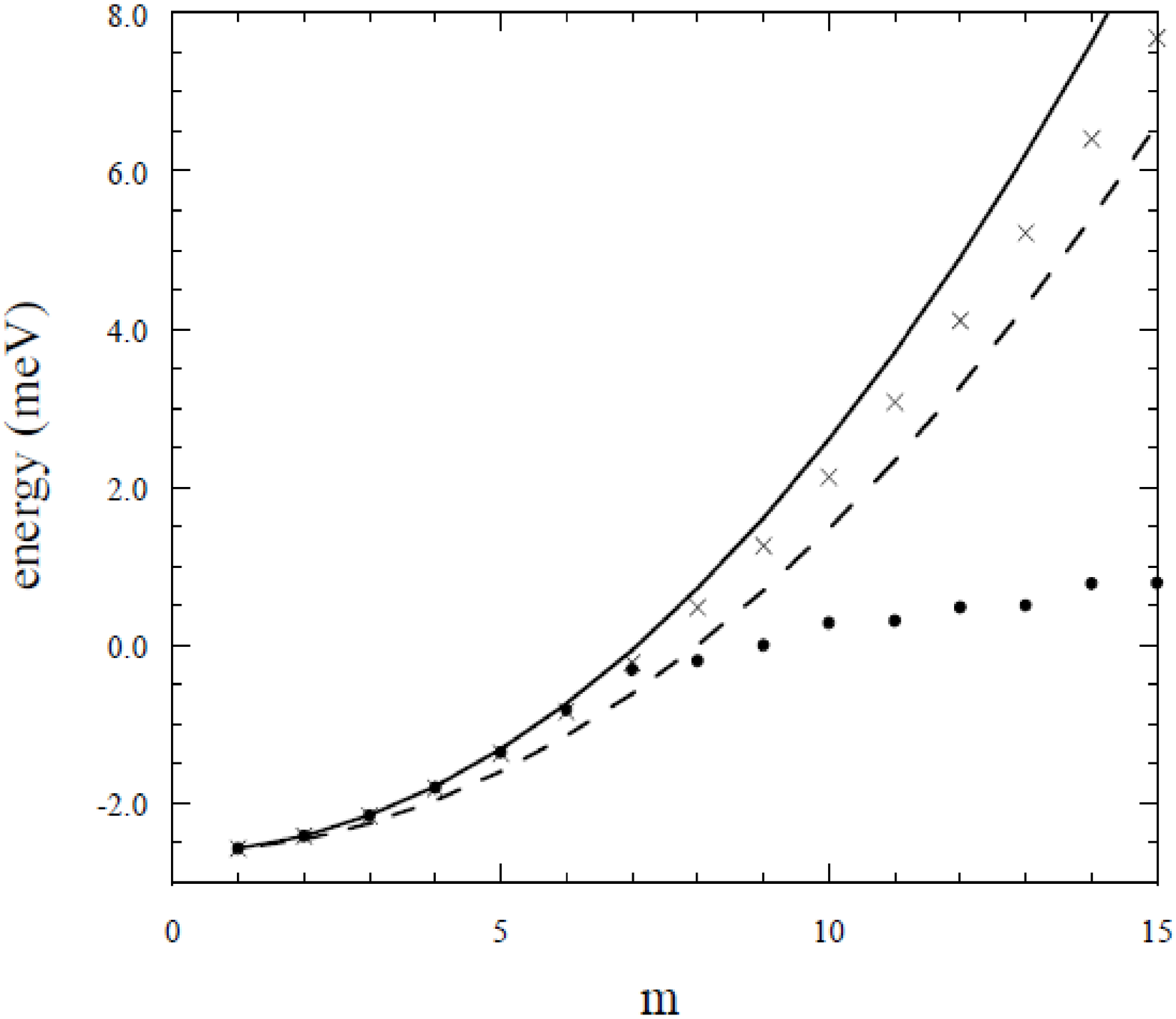}
\caption{Light hole exciton energy as a function of centre-of mass quantum number m in a AlGaAs/GaAs(001) quantum well of thickness L=250nm. The energy is computed by full theory (dots), analytical model (crosses), hard sphere model with (solid curve) and without dead layer (dashed curve).}
\end{figure}

In Fig.5 and 6 the heavy and light hole exciton energies, computed by the full theory of ref.[32] and by the former analytical model for a quantum well of thickness L=$20a_B $=250nm are given as a function of an energy order parameter m that in the case of our analytical model coincides with the centre-of-mass quantum number. The physical parameters adopted in the calculations are the same of ref. [36], and the minimized parameters are: $a_{lh} = 19.977 \pm 0.002nm$ and $a_{hh} = 12.483 \pm 0.002nm$ for light and heavy hole effective Bohr radii respectively, and $\delta _{lh} = 14.277 \pm 0.002nm$, $\delta _{hh} = 21.133 \pm 0.002nm$ for the corresponding transition layers.

We observe, from Fig.5 and Fig.6, that the contribution of n=2 hydrogenic function becomes important for heavy hole exciton states with the centre-of-mass quantum number greater than m$>$12 corresponding to the exciton energy $\sim 2.3meV$ greater than the heavy hole  m=1 state, and analogously for light hole states m$>$8 ($\sim 3.8meV$). Therefore, the analytical model, also in its lowest order of formulation\cite{dads1}, should be able to reproduce the optical response in this range of quantum well dimensions since both the former ranges of energy are greater than those shown in the experimental spectra (see Fig.14b) of Schneider et al.\cite{Schneider}.

Finally, in Fig.5 and Fig.6 also the "hard sphere" model energies, computed with and without ($d \to 0$) the dead layer effect, are shown. The rather sound agreement between the full theory and the hard sphere model with dead-layer effect, for the lower m-values is essentially due to the values adopted in the present calculation, that coincide with the transition layer values obtained by a variational minimization (namely: $\delta _{hh}  = 1/\bar P_{hh}  \approx $21.1nm and $\delta _{lh}  = 1/\bar P_{lh}  \approx $14.4nm) while those computed with the hard sphere interpolation equation: $d = \left( {1 - 2\mu /M} \right)\,a_B $ are strongly different from the former ones ($d_{hh}  \approx $ 10nm and $d_{lh}  \approx $6.5nm for heavy and light hole exciton respectively).

Now, let us compute the lowest energy envelope function of an exciton in the hydrogenic limit ($m_h  \to \infty $) for the heavy hole trapped at the site $Z_o  = 0$ of a rather large quantum well ($L = 20a_B $). The analytical envelope function is: $\Psi _m (z;Z_o  = 0) = N_m \,g_m (z;Z_o  = 0)\,\varphi _{1S} (r)$, and, for transition layer limit $\delta  \to 0$ ($\bar P \to \infty $), the exciton energy converges to the bulk value. Indeed, in this limit, from the eqs. (5a) and (5b) , we obtain $tg(K_m L/2) \to  - \infty $ and $tg(K_m L/2) = 0$ respectively, therefore the wave vector quantization of the electron is: $K_m  = \frac{\pi }{L}\,m$ for m=1,2,3,\textellipsis\,.
\begin{figure}[t]
\includegraphics[scale=0.3]{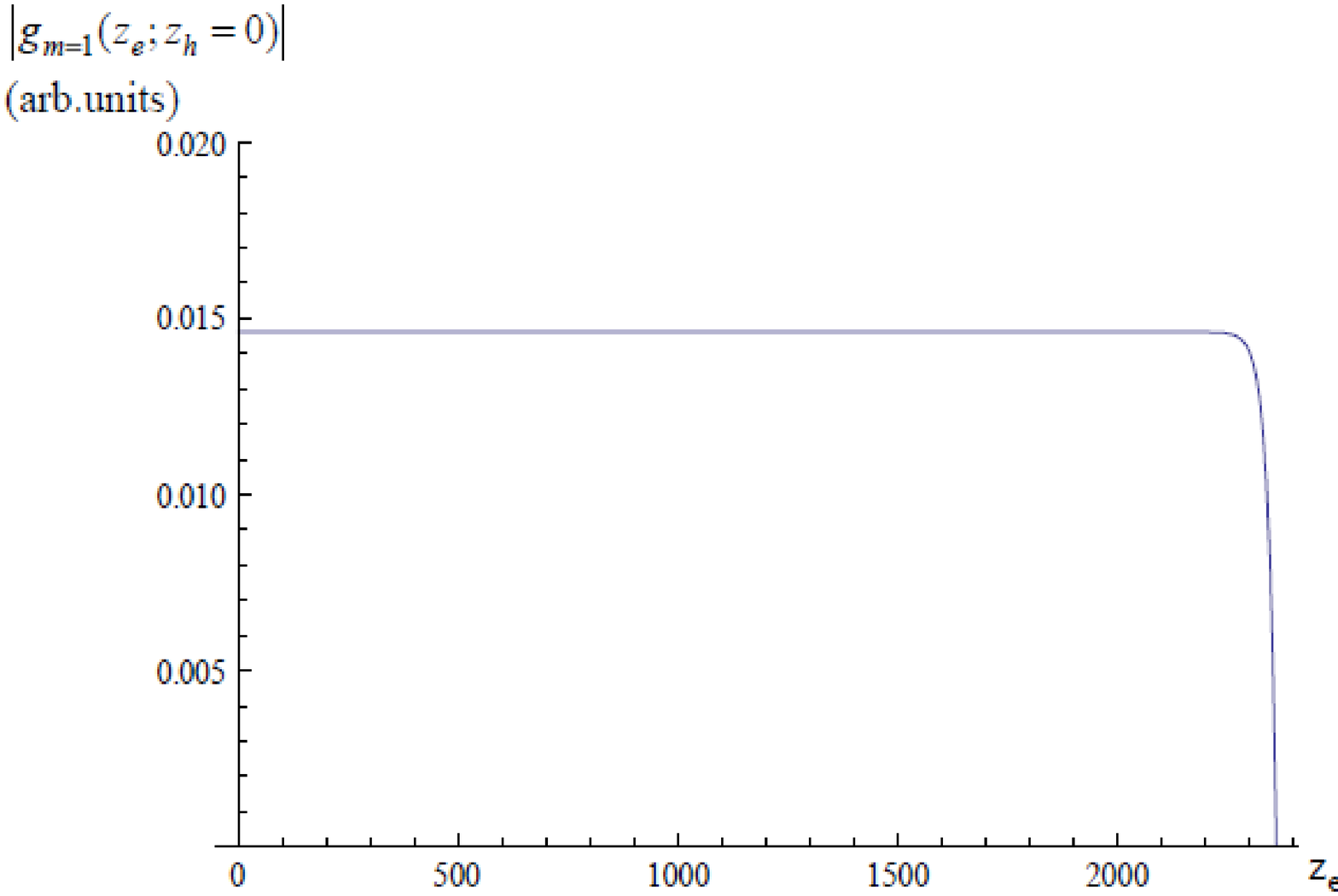}
\caption{Trapped exciton confinement function $\left| {g_{m = 1} (z_e ;z_h  = 0)} \right|$ (eq.(2B)) of  AlGaAs/GaAs(001) quantum well of thickness L=250nm as a function of electron Z-coordinate with transition layer$\delta  \to 0$.}
\end{figure}
\begin{figure}[t]
\includegraphics[scale=0.32]{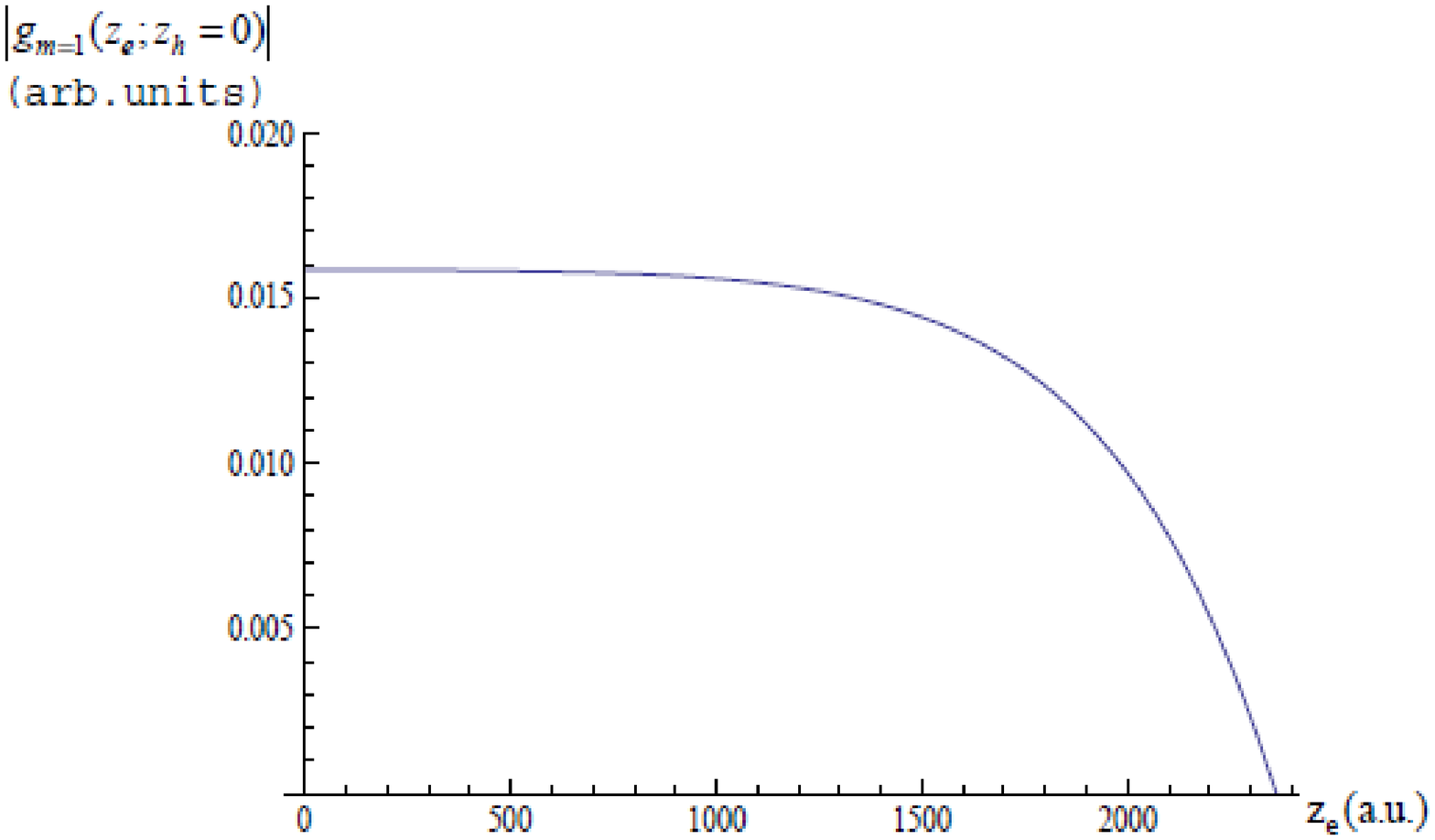}
\caption{Trapped exciton confinement function $\left| {g_{m = 1} (z_e ;z_h  = 0)} \right|$ of AlGaAs/GaAs(001) quantum well of thickness L=250nm as a function of electron Z-coordinate with transition layer $\delta = 21.16nm$.}
\end{figure}
In appendix B the first momentum of the lowest energy state of the trapped exciton Hamiltonian in a large quantum well is given in an analytical form. The bulk exciton energy can be obtained also by a variational minimization as a function of effective Bohr radius $a = a_B  = 10.8711 \pm 0.002nm$ and the effective Rydberg is Ryd=-4.81173meV. In Fig.7, the confinement function ${g_1 (z;Z_o  = 0)}$ is computed as a function of the electron-hole relative motion along z-axis.  Notice, that, at variance of a simple cosine function that describes electron confinement in a 2D quantum well\cite{Combescot17}, this function shows a rather flat shape, except at the interfaces ($Z \approx  \pm L/2$), and, as fully discussed in ref. [27,28], this behaviour is a fingerprint that the present analytical model is well suited for describing the centre-of-mass quantization.
In Fig.8 we compute the former system by imposing a transition layer value different from zero ($\delta$ =21.133nm); also in this case the shape remains rather smooth, but the transition layer effect is clearly shown at the boundaries of the well.

In conclusion we would like to remind that the optical response of exciton-polaritons is strongly dependent, not only on the exciton energy, but also on its envelope function shape. Therefore for very large quantum wells ($L\geq 20a_B$) the former analytical model should be able to describe the optical experimental mesurements of ref.[2].

In order to generalize the former model, also in the zone very close to the quantum well thicknesses $5a_B  \leqslant L \leqslant 10a_B $, we will have to extend the finite sum of eq.(3) to higher energies of the hydrogenic wave function, as discussed in the appendix A. Moreover, since the real symmetric matrix Hamiltonian is rather diagonal only for large wells, in this range of thicknesses, we must solve a generalized eigenvalue problem, namely: $\overset{\lower0.5em\hbox{$\smash{\scriptscriptstyle\leftrightarrow}$}} {A} \;\vec \Psi  = E\;\overset{\lower0.5em\hbox{$\smash{\scriptscriptstyle\leftrightarrow}$}} {B} \;\vec \Psi $
where $\left\langle {\Psi _n } \right|\hat H_{ex} \left| {\Psi _m } \right\rangle  = \left\langle {\Psi _m } \right|\hat H_{ex} \left| {\Psi _n } \right\rangle  = \overset{\lower0.5em\hbox{$\smash{\scriptscriptstyle\leftrightarrow}$}} {A} $
and the matrix of the normalized exciton states is quasi-unit real symmetric matrix $\left\langle {\Psi _n } \right|\left| {\Psi _m } \right\rangle  = \left\langle {\Psi _m } \right|\left| {\Psi _n } \right\rangle  = \overset{\lower0.5em\hbox{$\smash{\scriptscriptstyle\leftrightarrow}$}} {B} $. Notice, that the generalized analytical envelope function is not a pure state of the center-of-mass wave vector, even if it still remains under the one-transition layer approximation.

In conclusion, we have pointed out that the analytical model is essentially based on a smart truncation of the hydrogenic basis set expansion, and it must be implemented (see appendix A) to correctly take into account the Wannier exciton dynamics in all the range of thickness till the semi-infinite limit ($5a_B  \leqslant L \leqslant \infty $).

\section{Optical response of a single quantum well under the center-of-mass quantization.}
In this section, the exciton-polariton propagation in a single quantum well under centre-of-mass quantization is computed by the former analytical exciton model in a large range of quantum well thicknesses ($5a_B  \le L \le 20a_B $
), and compared with the full theories of refs. [2,31,36], and with the heuristic hard sphere model. Moreover, we will compute self-consistently the polariton propagation in a slab in order to reproduce, without introducing fitting parameters (except the homogeneous non-radiative broadening), the experimental transmission spectra in the high quality GaAs samples of Schneider et al.\cite{Schneider}

First of all, let us consider a GaAs quantum well of thickness $L = 5a_B =62.5nm$, that is in the transition zone, as discussed in the former section. The parameter values, adopted for the calculation, are taken from ref.[36], namely: $m_e  = 0.067m_o $, $m_{hh}  = 0.457m_o $, background refraction index $n_b  = 3.71$, energy gap $E_{gap}  = 1.42eV$, and non-radiative broadening $\Gamma$=0.03meV. The electric dipole moment is replaced by the Kane's energy\cite{Kane} ($E_K = $22.71eV for GaAs), that is conceptually more suitable than the electric dipole moment for describing the radiation-matter interaction in translational periodic systems.

The transmittance spectrum is computed by a self-consistent solution of Schroedinger-Maxwell equations in the effective mass approximation, as reported in ref.[31], where all the analytical formula necessary for the calculation are given explicitly. The Fabry-Perot effect is suppressed in the spectra\cite{Schneider,Schumacher} by using the same background dielectric function in whole the space; this allows to observe exciton polariton peaks free from the interferences due to the Fabry-Perot oscillations\cite{Schneider}.
\begin{figure}[t]
\includegraphics[scale=0.3]{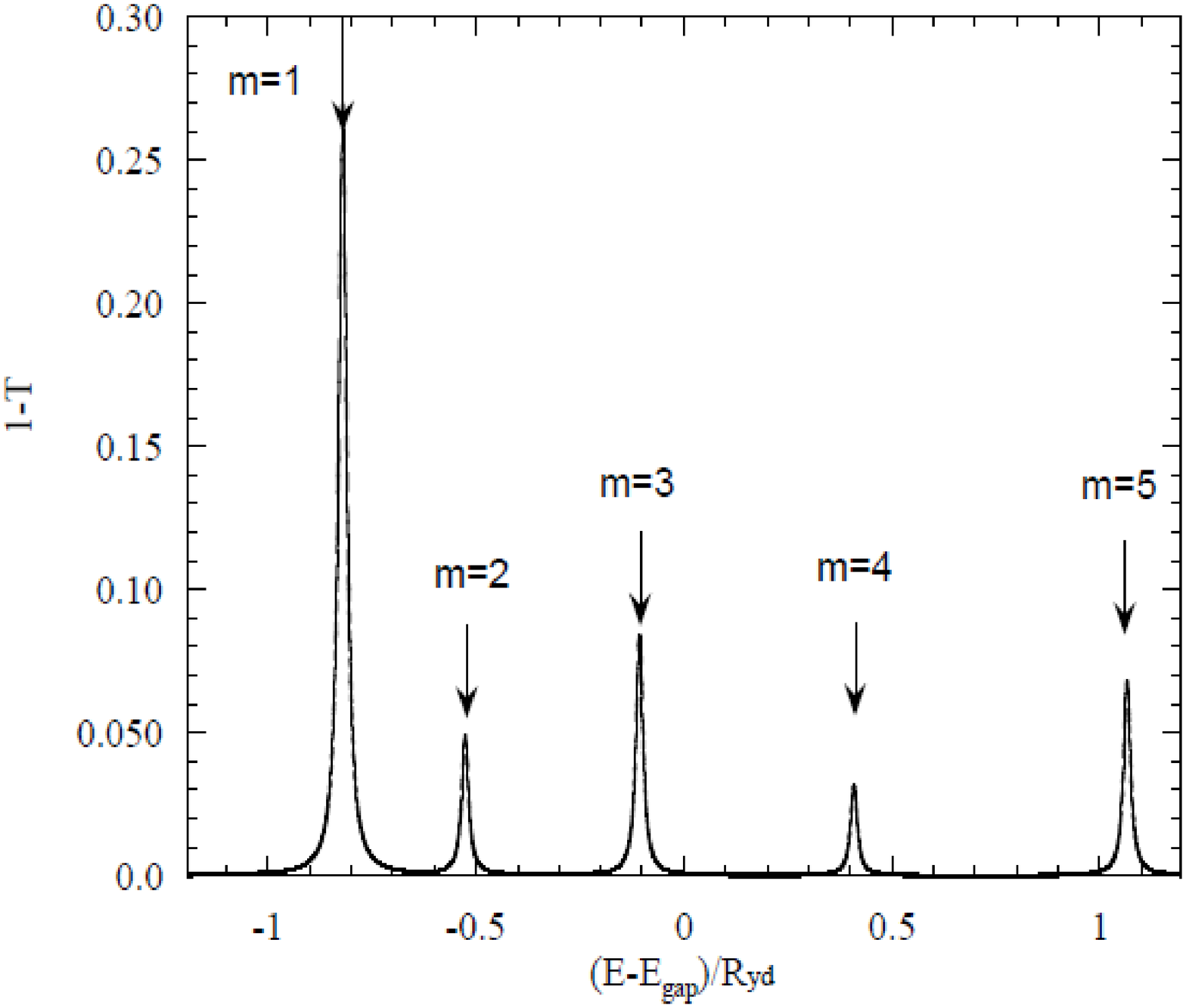}
\caption{Optical absorbance (1-T) of the heavy hole exciton as a function of photon energy ($(E - E_{gap} )/Ryd$) in  AlGaAs/GaAs(001) quantum well of thickness L=62.5nm.}
\end{figure}
\begin{figure}[b]
\includegraphics[scale=0.3]{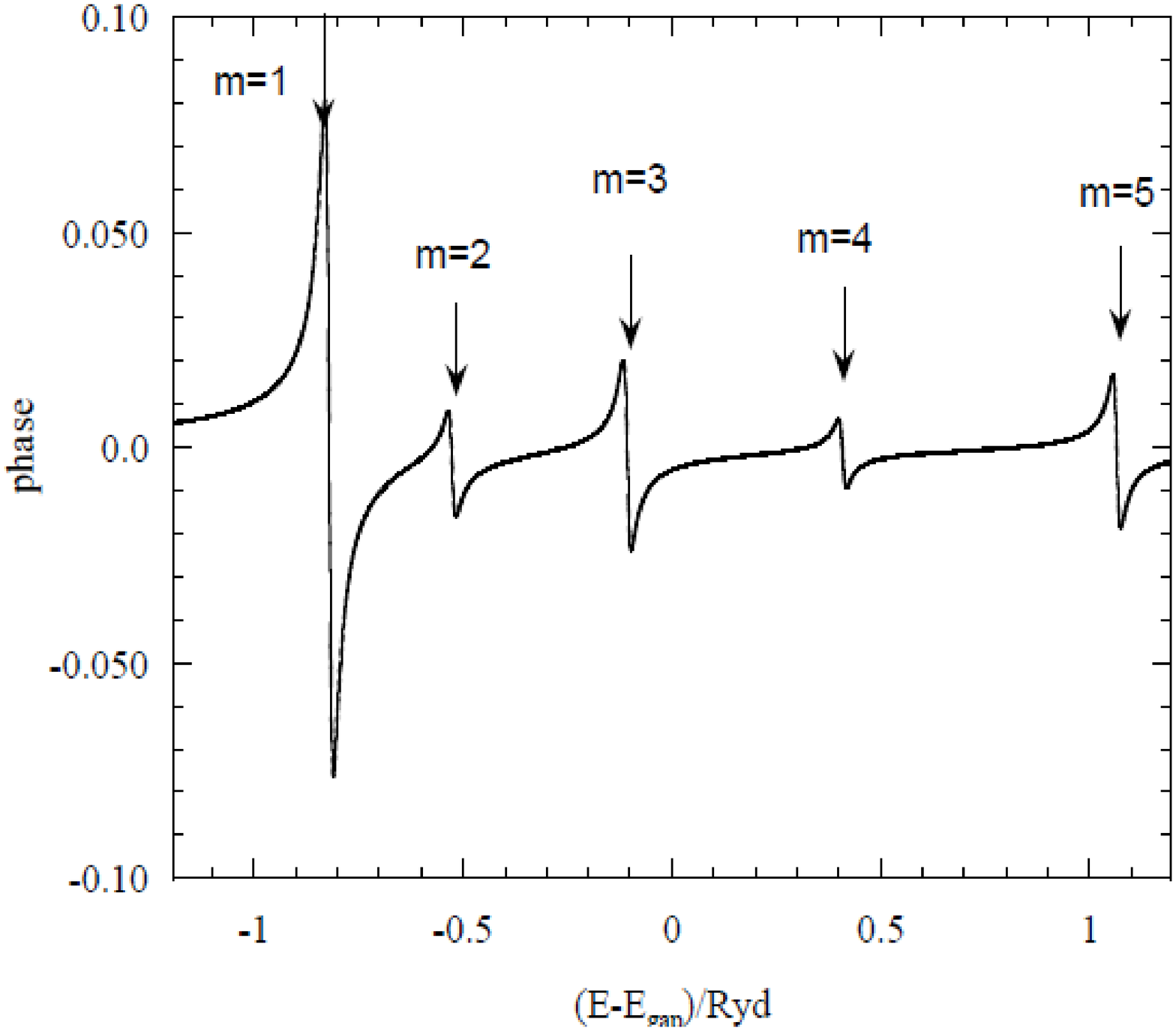}
\caption{Phase of the optical absorbance of the heavy hole exciton as a function of photon energy ($(E - E_{gap} )/Ryd$) in AlGaAs/GaAs(001) quantum well of thickness L=62.5nm.}
\end{figure}
In Fig.9 the absorption spectrum is shown for the lowest five polariton states; the energy positions of the optical peaks are labelled with the same quantum number m of the corresponding centre-of-mass exciton state (m=1$ \div $5). For this quantum well thickness (L=$\lambda/4$) the even and odd exciton-polariton states show almost comparable intensities. 

The phase of the transmittance amplitude is shown in Fig.10. Notice that the maxima of the absorption are in correspondence of the inflection points of the phase curve; therefore, there is a complete correspondence in the polariton spectra between phase and intensity. To properly reproduce both these quantities is a very severe check for the exciton model adopted in the calculation as fully discussed in ref. [2] and underlined in the present theory versus the experimental results.
\begin{figure}[t]
\includegraphics[scale=0.3]{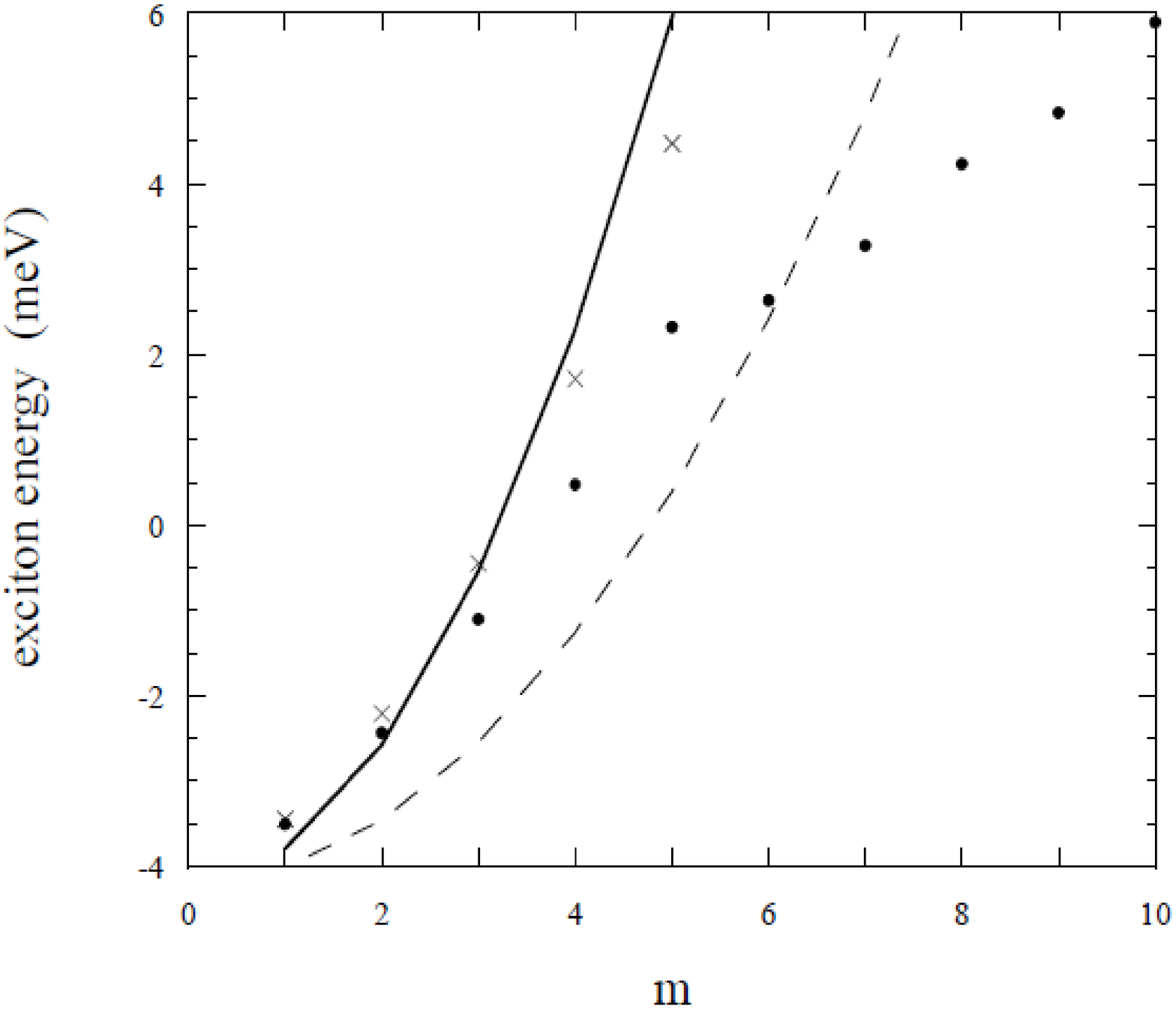}
\caption{Heavy hole exciton energy as a function of centre-of-mass quantum number m in  AlGaAs/GaAs(001) quantum well of thickness L=62.5nm. The energy is computed by full theory (dots), analytical model (crosses), hard sphere model with (solid curve) and without dead layer (dashed curve).}
\end{figure}

Now, let us compare the exciton energies computed by the analytical model with those derived by the full theory of ref. [32]. In Fig.11 these exciton energies are shown as a function of the quantum number m of the centre-of-mass motion. The minimized effective Bohr radius and transition layer of the analytical model are: a=12.464$ \pm $0.002nm and $\delta$=$1/\bar P$=10.228$ \pm $0.002nm respectively. The exciton energy computed with the simple hard sphere model are also reported for vanishing dead layer and for a dead layer as large as the minimized transition layer value. We note a rather large energy difference  (about 1,23meV for m=4) between analytical model and full theory in correspondence with higher energy states (m$ \ge $4), that rapidly decreases for the lower energy states; for instance for m=3 it drops to 0.6meV and for m=2 is about 0.2meV. The last value is rather acceptable if we take into account that the non-radiative broadening is about $\Gamma _{NR}  = 0.5 \div 1.0\,meV$ in tipycal samples and $\Gamma _{NR}  \approx 0.25meV$ in high quality quantum wells. Obviously, the large discrepancy observed for the high energy states is not surprising if we take into account that the exciton envelope functions of the analytical model, labelled with the quantum number m, are pure states of the centre-of-mass, while the exciton states of the full theory are a general superposition of states with different centre-of-mass wave vectors and are expanded on a rather complete set of the relative motion wave functions.

In conclusion, in this thicknesses range of values the lowest exciton states are in sound agreement with those obtained by the full theory, while for describing higher energy states of the centre-of-mass (m$ \ge $4) we will have to implement the simple analytical model as discussed at the end of the former section.  Moreover, we observe that the full theory gives exciton states located between the two curves of the hard sphere model with and without the dead layer effect and this leads to the well known possibility of recovering the rather correct exciton energies by using the Pekar's dead-layer as a fitting parameter.
\begin{figure}[t]
\includegraphics[scale=0.3]{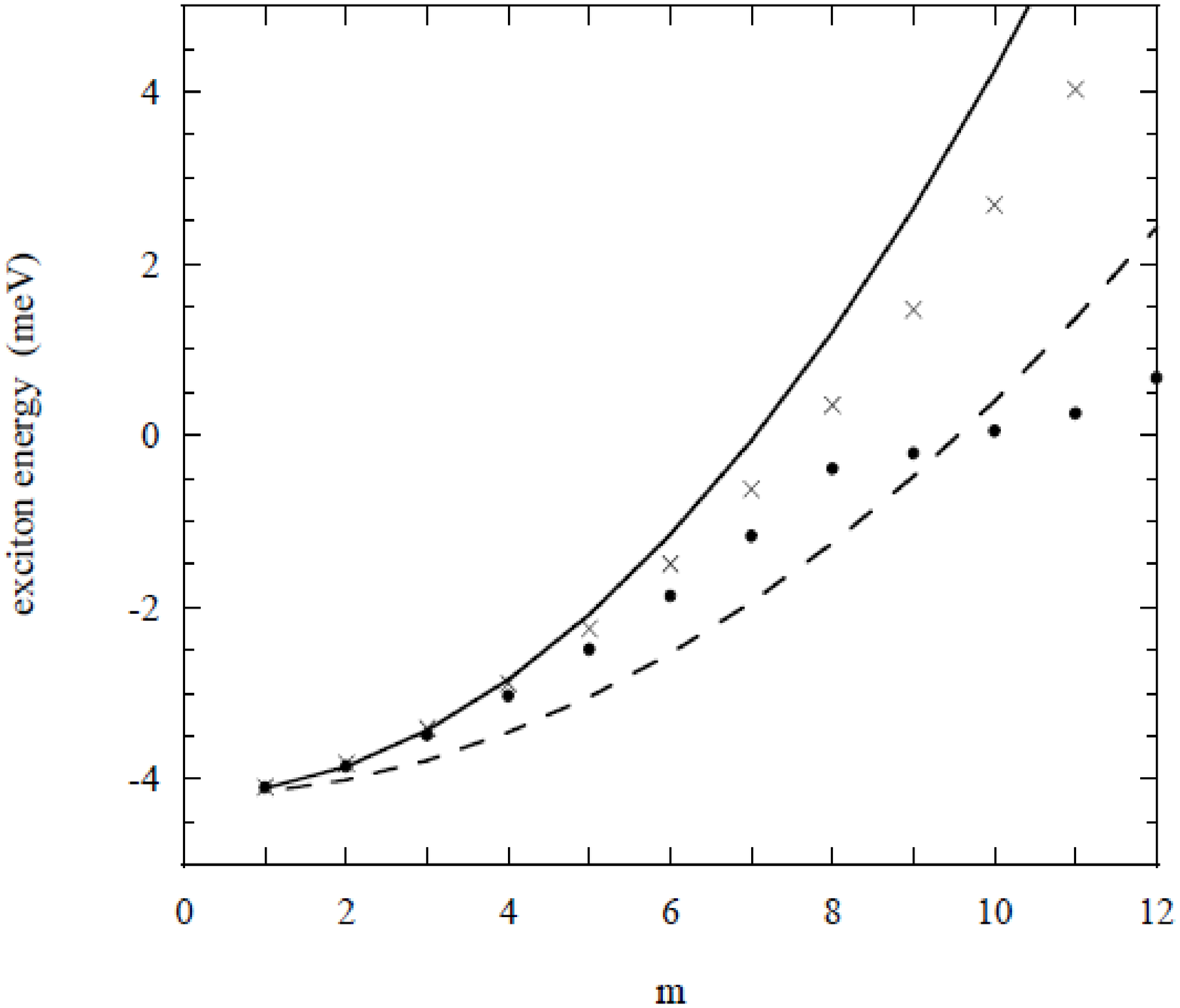}
\caption{Heavy hole exciton energy as a function of centre-of mass quantum number m in  AlGaAs/GaAs(001) quantum well of thickness L=125nm. The energy is computed by full theory (dots), analytical model (crosses) , hard sphere model with (solid curve) and without dead layer (dashed curve).}
\end{figure}
\begin{figure}[b]
\includegraphics[scale=0.9]{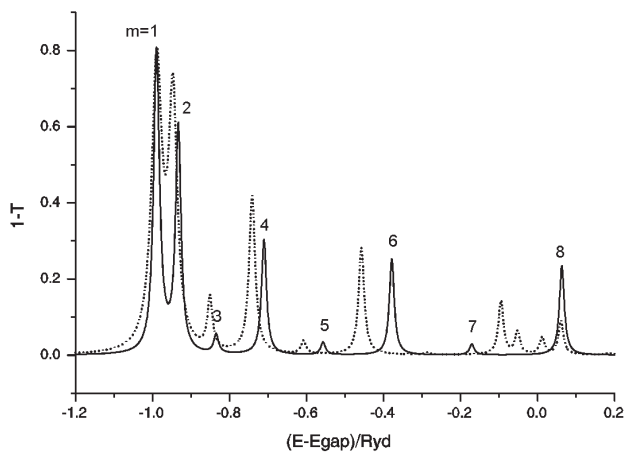}
\caption{Optical absorbance (1-T) of the heavy hole exciton as a function of photon energy ($(E - E_{gap} )/Ryd$) in  AlGaAs/GaAs(001) quantum well of thickness L=125nm: full theory (dashed curve), analytical model (solid curve).}
\end{figure}
Now, let us consider GaAs quantum wells in the zone of the centre-of-mass quantization $10a_B  \le L \le 20a_B $. In Fig.12 the exciton energies of the analytical model and of the full theory\cite{atanasov} in a single quantum well of  thickness L=$10a_B $=125nm are shown as a function of the centre-of-mass quantum number m. The physical parameters values are the same of the former sample, and the minimized variational parameters are: $a = 12.560 \pm 0.002nm$ and $\delta  = 1/P = 16.400 \pm 0.002nm$. We observe that, while the effective Bohr radius is very close to the 3D Bohr radius, the transition layer value is rather different from its saturation value, underlining the strong influence of the transition layer effect on the exciton dynamics in this range of  thicknesses. The calculation confirms a neglegible difference between the two models for the lower energy states ($m \leqslant 4$), and an increase of this difference till 0.75meV for the m=8 state where higher energy ($n>1$) hydrogenic states cannot be neglected. In fact, exciton states with m$ \ge $8 drop into the second parabolic curve due to the n=2 hydrogenic states, therefore the agreement between the analytical model and the full theory can be improved along the line discussed in the appendix A. In Fig.12 also the exciton energies computed by Pekar's model with and without dead-layer are shown for sake of comparison.

In Fig.13 the optical spectra, computed self-consistently with the analytical exciton envelope function and with the full exciton model, are shown; due to the photon wavelength that is about $\lambda  \approx 2L$ the odd exciton functions (m=2,4,6,\textellipsis) give a stronger contribution than the even (m=1,3,7,\textellipsis) ones. In fact, in the computed spectrum of the full exciton theory\cite{atanasov} the intensity of m=7 exciton-polariton peak is vanishingly small, while it is shifted towards the higher energies for the analytical model. We would like to underline that the transmission spectrum computed by our full theory is in very good agreement, both in energy positions and in line-shapes, also with those computed by the full theories of  refs. [2] and [36], not reported here.

In order to go a bit deeper in the optical spectrum analysis, notice that the double peak, due to the lowest exciton energies (m=1,2), is very sensitive to the radiation-matter interaction.
Moreover, due to the small non-radiative broadening value adopted in the calculation, we can appreciate the difference in energy between the analytical model and the full theory for all the range of energies of  the polariton spectrum even if theirs values remain in any case rather small for $m < 8$
exciton-polariton states. In fact, let us consider the two peaks for m=6: they seem very well separated in the spectrum but the energy difference ($<$0.4meV) is lower than the optical broadening (see also Fig.12).

Finally, we would like to underline that in the range of energies reported in Fig.13 our microscopic model gives a more accurate result than that based on Pekar's ABC. In fact, in ref. [36] (see fig.4) the energy difference between full theory and Pekar's model for m=6 exciton-polariton state is about 0.7 meV, that is two-times our difference, and it is greater than the broadening value. Therefore, we can conclude that our analytical model can reproduce the full theory till $m \le 6$ exciton-polariton state, and that, in any case, it gives a better agreement with the full theories than that obtained by Pekar's model\cite{Schumacher}.
\begin{figure}[b]
\includegraphics[scale=0.5]{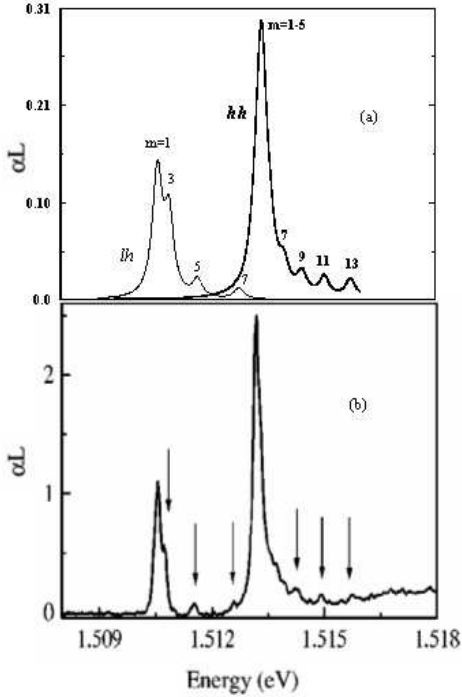}
\caption{(a) Optical absorbance (1-T) computed for light  and heavy hole exciton as a function of  photon energy in  AlGaAs/GaAs(001) quantum well of thickness L=250nm. (b) Experimental optical absorbance of light and heavy hole exciton of ref. [2] as a function of photon energy in  AlGaAs/GaAs(001) quantum well of thickness L=250nm.}
\end{figure}

Now, let us try to reproduce the experimental transmission spectrum of an high quality GaAs quantum well reported in the work of Schneider and al.\cite{Schneider} by our analytical self-consistent theory. In fig.7b of ref. [2], reproduced here in Fig.14b, many peaks due to the centre-of-mass quantization of both heavy and light hole excitons in a large quantum well of GaAs ($L = 20a_B  = 250nm$) are clearly shown by arrows. We would like to underline that we have choosen to reproduce the experimental results of Schneider et al.\cite{Schneider} because they are well suited for checking the present analytical model. In fact, the Wannier exciton is perfectly confined in a self-sustained quantum well, and the Fabry-Perot effect, due to the modulation of the background dielectric function, is suppressed by the presence of antireflection coating on both the sample surfaces. Moreover, in a large quantum well the anisotropic electron-hole masses induce a very small energy shift on the exciton energies, therefore, we can use isotropic exciton masses in the present range of  thicknesses.  

In the former section, we have shown that the analytical exciton model is in perfect agreement with the full theory\cite{atanasov} in a sufficient large range of energies (9meV), and this is well suited for reproducing the experimental spectrum of ref. [2]. The parameter values adopted for the calculation are\cite{Schneider}: $L = 20a_B  = 250nm$, $m_e  = 0.067$, $m_{hh}  = 0.457m_o $, $m_{lh}  = 0.080m_o $, $E_{gap}^{lh}= 1.5105eV$, $n_b  = 3.71$, $E_{gap}^{hh}= 1.5130eV$ and Kane's energy $E_K  = 22.71eV$; the minimized parameter values, given in the former section (see also Fig.5 and Fig.6), underline the different behaviours, from positronium for light hole to hydrogenic atom for heavy hole, of Wannier excitons. The energy shift between heavy and light hole energy gaps, due to the residual strain of the sample, is discussed and evaluated in ref.[2].

The high quality ($\Gamma _{NR}  \leqslant 0.25meV$) of  the samples should allow to study the effect induced by the wave vector quantization of the centre-of-mass motion at T=2K. Indeed, the polariton effect is important only in the case where the exciton broadening value is less than a critical value\cite{Tredicucci}, namely: $\Gamma _c  = \left[ {\frac{{8\Delta _{LT} E_{ex}^2 \varepsilon _b }}{{Mc^2 }}} \right]^{1/2} $, that for the parameters chosen is $\Gamma _c  \approx 0.21meV$. Notice that the non-radiative broadening value chosen in the present calculation, in order to reproduce the experimental spectra lineshapes ($\Gamma _{NR}  = 0.15meV$), is in rather close agreement with that given in ref.[2], and, moreover, since $\Gamma  < \Gamma _c $, exciton-photon coupling becomes dominant with respect to the exciton acoustic phonon interaction.

Since the exciton is computed in a simple two bands model, the optical response of  heavy and light hole exciton-polaritons will be shown in the pictures by different curves but in the same energy scale.

The absorbtion of light and heavy hole excitons are shown in Fig.14a. We would like to underline that the agreement between theory and experiment is very good both in the line shapes and in energy positions, as shown by the arrows in correspondence of the maxima of the experimental spectrum of Schneider\cite{Schneider} reported in Fig.14b. In fact, the characteristic double peak of the light hole exciton-polariton and the asymmetric line shape of the main peak of the heavy hole, due to the interplay of polaritonic effect and centre-of-mass quantization of the Wannier exciton, are clearly reproduced.
This agreement is particularly meaningful when we take into account that it is obtained for the same parameter values in the presence of rather different exciton dynamics. In fact, light hole exciton is close to positronium limit ($\mu /M$=0.248), while the heavy hole is rather close to the hydrogenic behaviour ($\mu /M$=0.111). Moreover, the analytical model allows to unambiguously assign the center-of-mass character for any features present in the experimental spectrum. For instance, the peak at m=7, not assigned in ref.[2], has even parity character, due to the photon wavelength ($L \approx \lambda $) that gives odd exciton-polariton peaks neglegible small in the spectrum.

\begin{figure}[t]
\includegraphics[scale=0.5]{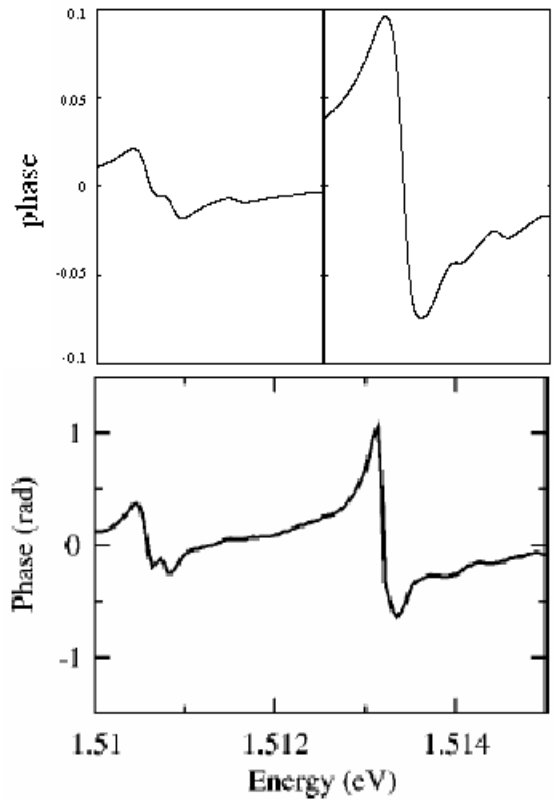}
\caption{(a) Phase of the optical absorption computed for light and heavy hole exciton as a function of photon energy in  AlGaAs/GaAs(001) quantum well of thickness L=250nm.  (b) Measured phase of the optical absorption\cite{Schneider} of light and heavy hole exciton as a function of photon energy in AlGaAs/GaAs(001) quantum well of thickness L=250nm.}
\end{figure}

Finally, notice that for the heavy-hole exciton polariton in ref.[2] (see Fig.10c) the m=11 peak is shifted toward the lower energy side, with respect to the experimental spectrum, while the m=13 one is out of the energy scale. 

In Fig.15a and 15b the computed phase of exciton-polaritons absorbtion spectrum and the corresponding experimental results of ref.[2] are shown respectively. Also in this case the agreement between theory and experiment is sound both for energy positions and lineshapes. Obviously, the long tail shown by the heavy-hole exciton-polariton phase towards the lower photon energies can not be matched with the light hole phase by the simple two-bands model. Moreover, all the features observed in the absorption spectrum have a counterpart in the phase spectrum, and also the characteristic two peaks feature of the light hole exciton-polariton and the centre-of-mass peaks at higher energies are reproduced.

In conclusion, the simultaneous measurements of phase and transmission spectra allow to assign the exciton-polariton peaks of the experimental spectrum to the pure centre-of-mass exciton-polariton quantization.

We would like to underline that in ref. [2] the authors compare the optical properties, computed by their accurate microscopic exciton model, with a number of heuristic models, based on different additional boundary conditions (ABCs), in order to conclude that "A simultaneous description of the measured amplitude and phase of the transmitted electric field is only possible with a \textit{full} model". The sound agreement between our analytical microscopic model, based on the one-transition layer approximation, and the experimental results of ref.[2]  strongly suggest the replacement of the adjective \textit{full} with \textit{microscopic}.

\section{Conclusions}
In the present work the analytical microscopic exciton model in a large ($L \gg a_B $) single quantum well, proposed in ref. [1] is revised and implemented in order to reproduce the Wannier exciton centre-of-mass quantization in the range of quantum well thicknesses $5a_B  \leqslant L \leqslant \infty $.
 
The microscopic transition layer effect, observed at the well/barrier interfaces in a single quantum well, is discussed at variance of the classical dead-layer introduced by Pekar\cite{Pekar}, and compared with various full theoretical models\cite{Schneider,atanasov,Schumacher}. 

Moreover, this new formulation has allowed to reconcile the Wannier exciton dynamics in large quantum wells ($L \gg a_B $) and in semi-infinite semiconductor samples ($L \to \infty $) described by the same analytical microscopic model\cite{dads1,viri}. 

Finally, exciton-polariton transmission spectra in large quantum wells ($L=20a_B $) are computed, and compared with experimental results in high quality GaAs samples obtained by Schneider et al.\cite{Schneider}. The sound agreement between theory and experiment has allowed to unambiguously assign the transmission spectra peaks to the pure states of the exciton centre-of-mass quantization.

\section*{Acknowledgments}
The authors are indebted to Mrs. E. Vuzza for technical support during the manuscript preparation.

\appendix
\section{A generalized variational Wannier exciton envelope function}
We generalize the Wannier exciton envelope function expanding eq.(2) of the text by taking into account also the 2s hydrogenic function. Since 2s is symmetric with respect to the z-axis we can  adopt the same procedure of Section II based on the following three steps.

1)	From the relationship $\hat{\mathbb{R}}\Psi _E (\vec r,Z) =  \pm \Psi _E (\vec r,Z)$ we obtain the even and odd envelope functions, namely:
\begin{subequations}
\begin{align}
\Psi _E^e (\vec r,Z)\, =\alpha _1 \Phi _E^e (\vec r,Z) + \hspace{8.in}\nonumber\\
2\sum\limits_{nlm}^{n > 2} {\left[ {\alpha _{nlm} \cosh \left( {P_n Z} \right)\varphi _{_{nlm} }^e (\vec r) - \beta _{nlm} \sinh \left( {P_n Z} \right)\varphi _{_{nlm} }^o (\vec r)} \right]} 
\hspace{6.in}\\
\nonumber\\  
\Psi _E^o (\vec r,Z)\, = \beta _1 \Phi _E^o (\vec r,Z) + 
\hspace{8.in}\nonumber\\
2\sum\limits_{nlm}^{n > 2} {\left[ {\alpha _{nlm} \cosh \left( {P_n Z} \right)\varphi _{_{nlm} }^o (\vec r) - \beta _{nlm} \sinh \left( {P_n Z} \right)\varphi _{_{nlm} }^e (\vec r)} \right]} 
\hspace{6.in}\nonumber\\
\end{align}  
\end{subequations}
where $\alpha _1 = \alpha_{100} $, $\beta_1 = \beta_{100}$; the functions $\Phi _E^i (\vec r ,Z)$ for i=e,o are: 
\begin{subequations}
\begin{align}
\Phi _{_E }^e (\vec r,Z)&= \cos \left( {K_1 Z} \right)\varphi _{1s} (r) + \tilde \alpha _{200} \cos \left( {K_2 Z} \right)\varphi _{2s} (r)
\hspace{2.in}\\
\Phi _{_E }^o (\vec r,Z)&= \sin \left( {K_1 Z} \right)\varphi _{1s} (r) + \tilde \beta _{200} \sin \left( {K_2 Z} \right)\varphi _{2s} (r)
\hspace{2.in}
\end{align}  
\end{subequations}
where $\tilde \alpha _{200}  = \alpha _{200} /\alpha _1 $ and $\tilde \beta _{200}  = \beta _{200} /\beta _1 $. Indeed, the functions of eqs.(1a.A) and (1b.A)  will be reduced to eqs.(4a.A) and (4b.A) respectively for $\alpha _{200}  = \beta _{200}  = 0$.

By adopting the one transition layer approximation an average value for the inverse of the transition layer depth, $P_{n > 2}= P = 1/\delta $ is taken and the envelope functions assume the following analytical form:
\begin{subequations}
\begin{align}
\Psi _E^e (\vec r,Z)=\hspace{9.in}\nonumber\\
=N_e \left[ {\Phi _E^e (\vec r,Z) + \cosh (PZ)F_{ee} (\vec r) - \sinh \left( {PZ} \right)F_{eo} (\vec r)} \right]
\hspace{6.2in}\\
\Psi _E^o (\vec r,Z) =\hspace{9.in}\nonumber\\
= N_o \left[ {\Phi _E^o (\vec r,Z) + \cosh (PZ)F_{oo} (\vec r) - \sinh \left( {PZ} \right)F_{oe} (\vec r)} \right]
\hspace{6.2in}
\end{align}  
\end{subequations}
where $N_e  = \alpha _1 $ and $N_o  = \beta _1 $.

2) We impose the no-escape boundary conditions (NEBCs): $\Psi _E (z_e=\pm L/2,z_h ,\vec \rho) = \Psi _E (z_h  =  \pm L/2,z_e ,\vec \rho ) = 0.0$
in order to compute the F-functions. 
For the even solution $\Psi _E^e (\vec r,Z)$, eq.(A3a), we obtain:
\begin{subequations}
\begin{align}
F_{ee} (\vec r) = \,f_{ee} (K_1 )\,\varphi _{1s} (\vec r) + \tilde \alpha _2 \,f_{ee} (K_2 )\,\varphi _{2s} (\vec r)
\hspace{2.in}\\
F_{eo} (\vec r) = \,f_{eo} (K_1 )\,\varphi _{1s} (\vec r) + \tilde \alpha _2 \,f_{eo} (K_2 )\,\varphi _{2s} (\vec r)
\hspace{2.in}
\end{align}  
\end{subequations}
for $0 \le z \le L/2$, $Z_1 (z) = \alpha _e z - L/2$ and $Z_2 (z) =  - \alpha _h z + L/2$. The f-functions for even exciton functions are:
\begin{subequations}
\begin{align}
f_{ee} (K) = \frac{{\cos (KZ_2 )\sinh (PZ_1 ) - \cos (KZ_1 )\sinh (PZ_2 )}}{{\sinh \left[ {P(Z_2  - Z_1 )} \right]}}
\hspace{2.in}\\
f_{oe} (K) = \frac{{\sin (KZ_2 )\cosh (PZ_1 ) - \sin (KZ_1 )\cosh (PZ_2 )}}{{\sinh \left[ {P(Z_2  - Z_1 )} \right]}}
\hspace{2.in}
\end{align}  
\end{subequations}
and analogously for the odd ones:
\begin{subequations}
\begin{align}
f_{oo} (K) = \frac{{\sin (KZ_2 )\sinh (PZ_1 ) - \sin (KZ_1 )\sinh (PZ_2 )}}{{\sinh \left[ {P(Z_2  - Z_1 )} \right]}}
\hspace{2.in}\\
f_{eo} (K) = \frac{{\cos (KZ_2 )\cosh (PZ_1 ) - \cos (KZ_1 )\cosh (PZ_2 )}}{{\sinh \left[ {P(Z_2  - Z_1 )} \right]}}
\hspace{2.in}
\end{align}  
\end{subequations}

3) From the continuity of the exciton wave function and its first derivative at the z=0 surface we obtain the conditions for the centre-of-mass quantization. Also in this case, due to the symmetry properties of the different components of the envelope functions, the centre-of-mass quantization must accomplish the two following relationships: $\left. {\frac{{\partial f_{ee} }}{{\partial z}}} \right|_{z = 0}  = 0$ and $\left. {\frac{{\partial f_{eo} }}{{\partial z}}} \right|_{z = 0}  = 0$ for the even exciton function (analogously for the odd one). It is simple to check that the centre-of-mass wave vector K  follows the even dispersion: $K_m^e \,tg\left[ {K_m^e \,L/2} \right] + P\,tgh\left[ {P\,L/2} \right] = 0$ for m=1,3,5,..., and the odd one: $P\,tg\left[ {K_m^o \,L/2} \right] - \,K_m^o \,tgh\left[ {P\,L/2} \right] = 0$ for m=2,4,6,.., according to the even ($f_{ee} (z)$,$f_{eo} (z)$) and odd ($f_{oo} (z)$,$f_{oe} (z)$) f-functions respectively. 
The generalized even (m=1,3,5,...) exciton function of eq.(A3a), assumes the analytical form:
\begin{eqnarray}
\Psi _{m_1 ,m_2 }^e (\vec r,Z) =\hspace{2.2in}\\
N_{m_1 ,m_2 }^e \left\{ {\cos (K_{m_1 } Z)\varphi _{1s} (\vec r) + } \right.\tilde \alpha _2 \cos (K_{m_2 } Z)\varphi _{2s} (\vec r) + \nonumber\\
 + \cosh (PZ)\left[ {f_{ee} (K_{m_1 } )\varphi _{1S} (r) + \tilde \alpha _2 f_{ee} (K_{m_2 } )\varphi _{2S} (r)} \right] + \nonumber\\
{ - \sinh (PZ)\left[ {f_{eo} (K_{m_1 } )\varphi _{1S} (r) + \tilde \alpha _2 f_{eo} (K_{m_2 } )\varphi _{2S} (r)} \right]\}}\nonumber
\end{eqnarray}
and analogously for the odd Wannier exciton function.

Notice that the inclusion in the analytical model of the 2s hydrogenic function not only increses the number of variational parameters, but also gives off-diagonal envelope function components in the centre-of-mass wave vector.

The inclusion in the model of 2p hydrogenic functions is a bit more tricky, since the finite sum is composed of two even components ($2p_x ,\,2p_y $), but also of an odd one ($2p_z $). Therefore, the generalized envelope functions should mix even and odd centre-of-mass wave vectors. In fact, let us choose ($\ell$,$
f$,$\zeta$) as the direction cosines of the relative motion coordinate with respect to the Cartesian axis: $\varphi _{2p} (\vec r) = \ell \,\varphi _{2p_x } (\vec r) + f\,\varphi _{2p_y } (\vec r) + \zeta \,\varphi _{2p_z } (\vec r)$.
In this case eqs. (A2a) and (A2b) become: 
\begin{subequations}
\begin{align}
\Phi _{_E }^e (\vec r,Z) = \cos \left( {K_{100} Z} \right)\varphi _{1s} (r) + \tilde \alpha _{200} \cos \left( {K_{200} Z} \right)\varphi _{2s} (r) + 
\hspace{2.1in}\nonumber\\
+\tilde \alpha _{211} \cos \left( {K_{211} Z} \right)\left[ {\ell \varphi _{2p_x } (\vec r) + f\varphi _{2p_y } (\vec r)} \right] +\hspace{3.in}\nonumber\\
+ \zeta \tilde \beta _{210} \sin \left( {K_{210} Z} \right)\varphi _{2p_z } (\vec r)\hspace{4.in}\\
\Phi _E^o (\vec r,Z)= \sin \left( {K_{100} Z} \right)\varphi _{1s} (r) + \tilde \beta _{200} \sin \left( {K_{200} Z} \right)\varphi _{2s} (r) + \hspace{2.1in}\nonumber\\
+\tilde \beta _{211} \sin \left( {K_{211} \,Z} \right)\left[ {\ell \varphi _{2p_x } (\vec r) + f\varphi _{2p_y } (\vec r)} \right] +\hspace{3.in}\nonumber\\
+ \zeta \tilde \alpha _{210} \cos \left( {K_{210} Z} \right)\varphi _{2p_z } (\vec r)\hspace{4.in}
\end{align}  
\end{subequations}
where $\tilde \alpha _{200}  = \alpha _{200} /\alpha _1 $, $\tilde \alpha _{211}  = \alpha _{211} /\alpha _1 $, $\tilde \beta _{210}  = \beta _{210} /\alpha _1 $
and $\tilde \beta _{200}  = \beta _{200} /\beta _1 $,
$\tilde \beta _{211}  = \beta _{211} /\beta _1 $, $\tilde \alpha _{210}  = \alpha _{210} /\beta _1 $. 
Now, let us impose the no-escape boundary conditions (NEBCs): $\Psi _E (z_e  =  \pm L/2,z_h ,\vec \rho ) = \Psi _E (z_h  =  \pm L/2,z_e ,\vec \rho ) = 0.0$ in order to compute the F-functions. 
For the even solution $\Psi _E^e (\vec r,Z)$  of eq.(A3a) we obtain:
\begin{subequations}
\begin{align}
F_{ee} (\vec r) = \,f_{ee} (K_{100} )\,\varphi _{1s} (\vec r) + \tilde \alpha _2 \,f_{ee} (K_{200} )\,\varphi _{2s} (\vec r) + 
\hspace{2.8in}\nonumber\\
 + \tilde \alpha _{211} f_{ee} (K_{211} )\left[ {\hat \ell \,\varphi _{2p_x } (\vec r) + \hat f\varphi _{2p_y } (\vec r)} \right] +\hspace{3.5in}\nonumber\\
 +\hat \zeta \,\tilde \beta _{210} f_{oo} (K_{210} )\varphi _{2p_z } (\vec r)\hspace{4.in}\\
F_{eo} (\vec r) = \,f_{eo} (K_{100} )\,\varphi _{1s} (\vec r) + \tilde \alpha _2 \,f_{eo} (K_{200} )\,\varphi _{2s} (\vec r) + 
\hspace{2.8in}\nonumber\\
 + \tilde \beta _{211} f_{oe} (K_{211} )\left[ {\hat \ell \,\varphi _{2p_x } (\vec r) + \hat f\varphi _{2p_y } (\vec r)} \right] +\hspace{3.5in}\nonumber\\
 +\hat \zeta \,\tilde \alpha _{210} f_{oe} (K_{210} )\varphi _{2p_z } (\vec r)\hspace{4.in}
\end{align}  
\end{subequations}
Finally, the generalized even (m=1,3,5,...) exciton function derivation proceed as before. Notice, that the former generalization leads to increase the number of variational parameters; in fact, while for n=1 we have two non-linear parameters (a,$\delta$), for n=2 we have two non-linear (a,$\delta$) plus three linear ($\tilde \alpha _{200} $, $\tilde \alpha _{211} $, $\tilde \beta _{211} $) parameters of minimization. 
Analogously for odd exciton state (m=2,4,6,...), and also in this case the solution embodies odd and even centre-of-mass wave vectors.
 
Finally, by adopting the former procedure, we obtain a set of normalized independent exciton states, and afterwards, it is necessary to solve the generalized problem in order to obtain an orthogonal set of states as discussed in Section II.

\section{Trapped exciton in a quantum well}
Let us consider a Wannier exciton trapped as a neutral atom at the site $\vec R_o  = (0,0,Z_o )$, whose electron mass is $m_e $, and the hole mass is taken in the limit $m_{hh}  \to \infty $. In this case the Z-component of the centre-of-mass coordinate is: $Z = z_h  = Z_o $, while the coordinate of the relative motion is: $z = z_e  - z_h  = z_e  - Z_o $; therefore, the motion of the electron along Z-axis is confined in the segment: $ - Z_o  - L/2 \le z \le  - Z_o  + L/2$. Since the envelope function of the trapped exciton has not defined parity, except for the site $Z_o  = 0$, a general variational trial function can be obtained as a linear superposition of even and odd exciton envelope functions in order to remove the restricted symmetry properties. Here, we adopt a more intuitive procedure observing that even and odd exciton envelope functions, defined in Sect.II, lost their symmetry property if projected into a non symmetric domain as the segment $ - Z_o  - L/2 \le z \le  - Z_o  + L/2$.  
In this case, m=1 envelope function, projected into the segment $ - Z_o  - L/2 \le z \le  - Z_o  + L/2$, can be used for a variational determination of the effective Bohr radius that should converges to the 3D Bohr radius for large quantum well ($L >  > a_B $), while the transition layer assumes the limit $\delta\to 0$. Obviously, many physical properties of exciton analytical model have to be redefined for trapped exciton. In fact, the projected envelope function is:
\begin{equation}
\Psi _m (\vec r;Z_o ) = N_m \,g_m (z;Z_o )\,\varphi _{1S} (r)
\end{equation}
For m odd (m=1,3,5,…) we have:
\begin{eqnarray}
g_m (z;Z_o ) =\cos (K_m Z_o ) + \hspace{1.6in}\nonumber\\
 +\cosh (PZ_o )f_{ee} (z) - \sinh (PZ_o )f_{eo} (z) \hspace{0.6in}
\end{eqnarray}
and for m even (m=2,4,6,..) :
\begin{eqnarray}
g_m (z;Z_o ) =\sin (K_m Z_o ) + \hspace{1.6in}\nonumber\\
 +\cosh (PZ_o )f_{oo} (z) - \sinh (PZ_o )f_{oe} (z) \hspace{0.6in}
\end{eqnarray}
where the f-functions are obtained by imposing the NEBCs at the surfaces of the well ($Z =  \pm L/2$). For $0 \le z \le L/2$ the confinement boundaries are $Z_1  =  - Z_o  - L/2 =  - L/2$ and $Z_2  =  - Z_o  + L/2 =  - z + L/2$:
\begin{eqnarray}
\tilde f_{ee} (z) = \left[ {\cos (KZ_1 )\sinh (PZ_2 ) - \cos (KZ_2 )\sinh (PZ_1 )} \right]\nonumber\\
\tilde f_{eo} (z) = \left[ {\cos (KZ_1 )\cosh (PZ_2 ) - \cos (KZ_2 )\cosh (PZ_1 )} \right]\nonumber\\
\end{eqnarray}
and for $ - L/2 \le z \le 0$ the confinement boundaries are $Z_1  = L/2$ and $Z_2  =  - z - L/2$:
\begin{eqnarray}
\tilde f_{oe} (z) = \left[ {\sin (KZ_1 )\cosh (PZ_2 ) - \sin (KZ_2 )\cosh (PZ_1 )} \right]\nonumber\\
\tilde f_{oo} (z) = \left[ {\sin (KZ_1 )\sinh (PZ_2 ) - \sin (KZ_2 )\sinh (PZ_1 )} \right]\nonumber\\
\end{eqnarray}
and $f_{ij} (z) = \tilde f_{ij} (z)/\sinh \left[ {P(Z_2  - Z_1 )} \right]$ for   i , j=e ,o .
Remember that the continuity of the exciton envelope function and its first derivative at the surface z=0 are the same of those given in eq.(5a) and (5b), but in the trapped exciton appears the electron quantization condition at variance of the former case that describes the center-of mass motion of Wannier exciton.
 
Taking into account that we have two different analytical equations for $z \ge 0$ and $z \le 0$, the normalization integration is:
\begin{eqnarray}
\left\langle {\Psi _m } \right|\left| {\Psi _m } \right\rangle  =\hspace{2.4in}\\
 2\pi \int\limits_0^\infty  {\rho d\rho \left\{ {\int\limits_{z_1 }^0 {dz}  + \int\limits_0^{z_2 } {dz} } \right\}} g_m^2 (z;Z_o )\varphi _{1S}^2 \left( {\sqrt {\rho ^2  + z^2 } } \right)\nonumber
\end{eqnarray}
and the first momentum of trapped exciton Hamiltonian:
\begin{equation*}
\hat H_{ex}^t  =  - \frac{{\hbar ^2 }}{{2\mu }}\vec \nabla _{\vec r}^2  - \frac{{e^2 }}{{\varepsilon _b r}}
\end{equation*}
is:
\begin{eqnarray}
\left\langle {\Psi _m } \right|\hat H_{ex}^t \left| {\Psi _m } \right\rangle  =  - \frac{{\hbar ^2 }}{{m_e }}\frac{1}{{a^2 }}\left\langle {\Psi _m } \right|\left| {\Psi _m } \right\rangle  +\hspace{1.in}\\
+2\pi \int\limits_0^\infty  {\rho d\rho \left\{ {\int\limits_{z_1 }^0 {dz}  + \int\limits_0^{z_2 } {dz} } \right\}} \left\{ {\frac{{\hbar ^2 }}{{m_e }}} \right.\left[ {\frac{1}{a} - \frac{1}{{a_B }}} \right]g_m^2 (z;Z_o ) + \nonumber\\
 - \frac{{\hbar ^2 g_m (z;Z_o ) }}{{2m_e }}\left[ {\left( {\left| z \right| + \frac{a}{2}} \right)\frac{{\partial ^2 g_m }}{{\partial z^2 }} - 2\frac{z}{a}\frac{{\partial g_m }}{{\partial z}}} \right]\varphi _{1S}^2 \left( {\sqrt {\rho ^2  + z^2 } } \right)\nonumber
\end{eqnarray}
The n-th derivative with respect to the relative motion along z-axis of the confinement functions $g_m (z;Z_o )$ are:
\begin{equation}
\frac{{\partial ^\ell  g_m }}{{\partial z^\ell  }} =  - \cos (PZ_o )\frac{{\partial ^\ell  f_{ij} }}{{\partial z^\ell  }} + \sin (PZ_o )\frac{{\partial ^\ell  f_{ij} }}{{\partial z^\ell  }}
\end{equation}
Taking into account that $dZ_2  =  - dz$, the derivative are:
\begin{equation}
\frac{{\partial f_{ij} }}{{\partial z}} = \left\{ {P f_{ij} \cosh \left[ {P\left( {Z_2  - Z_1 } \right)} \right] + \frac{{\partial \tilde f_{ij} }}{{\partial z}}} \right\}/\sinh \left[ {P\left( {Z_2  - Z_1 } \right)} \right]
\end{equation}
and
\begin{eqnarray}
\frac{{\partial ^2 f_{ij} }}{{\partial z^2 }} = - P^2 f_{ij} (z)+\hspace{2.3in}\\
 \left\{ {\frac{{\partial ^2 \tilde f_{ij} }}{{\partial z^2 }} + 2P\frac{{\partial f_{ij} }}{{\partial z}}\cosh \left[ {P\left( {Z_2  - Z_1 } \right)} \right] } \right\}/\sinh \left[ {P\left( {Z_2  - Z_1 } \right)} \right]\nonumber
\end{eqnarray}
Finally, for very large quantum wells the variational energy of trapped exciton recover the 3D Rydberg value, while the confinement function $\left| {g_m (z;Z_o )\,} \right|$ is an asymmetric function except for the hole coordinate $Z_o  = z_h  = 0$ and $ - L/2 \le z_e  \le L/2$ as discussed in the text (Fig.7 and 8).


\begin{references}
\bibitem{19}
Notice that the choice of $n \to \infty $ for N-1 degenerate virtual states, not strictly  necessary   for the consistency of the theory, gives two unpleasant results, namely: i) a too small transition layer value in the hydrogenic limit ($\mu /M \to 0$), as underlined in ref. [15], and ii) a physically inconsistent behaviour as a function of mass ratio $0 \leqslant \mu /M \leqslant 1/4$ (see M.Combescot in ref. [28]). Moreover in ref. [34,35] it was introduced also an extrinsic dead-layer in order to cure the first inconsistency, but not the second one. In the present paper a different choice, that does not show neither of these inconsistencies, is suggested.
\bibitem{25}
In order to perform very accurate variational minimization of the Wannier center-of -mass energy in a quantum well, we have used the equations given in ref. [31], where all the indeterminate (0/0) quantities,  present in the  implicit equations of ref. [1], were analytically eliminated in order to improve the numerical accuracy.
\end{references}
\end{document}